\documentclass[fleqn,10pt]{SelfArx}

\usepackage[english]{babel}
\usepackage{amsmath,amssymb,mathtools}
\usepackage{array}
\usepackage{listings}
\usepackage{xcolor}
\usepackage{orcidlink}
\usepackage{csquotes}
\usepackage{enumitem}
\usepackage[style=numeric,sorting=none,backend=biber]{biblatex}
\usepackage{hyperref}

\makeatletter
\DeclareFontFamily{OT1}{pcr}{\hyphenchar \font\m@ne}
\DeclareFontShape{OT1}{pcr}{m}{n}{<-> s * [0.90] pcrr7t}{}
\DeclareFontShape{OT1}{pcr}{m}{sc}{<-> s * [0.90] pcrrc7t}{}
\DeclareFontShape{OT1}{pcr}{m}{sl}{<-> s * [0.90] pcrro7t}{}
\DeclareFontShape{OT1}{pcr}{b}{n}{<-> s * [0.90] pcrb7t}{}
\DeclareFontShape{OT1}{pcr}{b}{sc}{<-> s * [0.90] pcrbc7t}{}
\DeclareFontShape{OT1}{pcr}{b}{sl}{<-> s * [0.90] pcrbo7t}{}
\DeclareFontShape{OT1}{pcr}{m}{it}{<->ssub * pcr/m/sl}{}
\DeclareFontShape{OT1}{pcr}{bx}{n}{<->ssub * pcr/b/n}{}
\DeclareFontShape{OT1}{pcr}{bx}{sc}{<->ssub * pcr/b/sc}{}
\DeclareFontShape{OT1}{pcr}{bx}{sl}{<->ssub * pcr/b/sl}{}
\DeclareFontShape{OT1}{pcr}{b}{it}{<->ssub * pcr/b/sl}{}
\DeclareFontShape{OT1}{pcr}{bx}{it}{<->ssub * pcr/b/it}{}
\DeclareFontShape{OT1}{pcr}{m}{ui}{<->ssub * pcr/m/it}{}
\DeclareFontShape{OT1}{pcr}{b}{ui}{<->ssub * pcr/b/it}{}
\DeclareFontShape{OT1}{pcr}{bx}{ui}{<->ssub * pcr/b/it}{}
\makeatother

\definecolor{color1}{RGB}{0,0,90}
\definecolor{color2}{RGB}{0,20,20}

\newcommand{\Lean}{\textsc{Lean}}
\newcommand{\mathlib}{\textsf{\small mathlib}}
\newcommand{\Fin}{\operatorname{Fin}}
\newcommand{\N}{\mathbb{N}}
\newcommand{\distH}{d_h}

\newcommand{\qary}{\texorpdfstring{$q$}{q}-ary}
\newcommand{\ArtifactRepoUrl}{https://github.com/florath/covering-codes-lean}
\newcommand{\ArtifactCommit}{460df105545c2d6b04ba71f29de6b56dbda92825}

\newcommand{\ArtifactBlob}{\ArtifactRepoUrl/blob/\ArtifactCommit}
\newcommand{\artifactrepo}{\href{\ArtifactRepoUrl}{\path{github.com/florath/covering-codes-lean}}}
\newcommand{\artifactcommit}{\href{\ArtifactRepoUrl/tree/\ArtifactCommit}{\texttt{\ArtifactCommit}}}
\newcommand{\leanfile}[1]{\href{\ArtifactBlob/#1}{\path{#1}}}
\newcommand{\leanfileat}[2]{\href{\ArtifactBlob/#1\#L#2}{\path{#1}}}
\newcommand{\leandir}[1]{\href{\ArtifactRepoUrl/tree/\ArtifactCommit/#1}{\path{#1}}}
\newcommand{\leansource}[3]{\href{\ArtifactBlob/#2\#L#3}{#1}}
\newcommand{\leanref}[3]{\href{\ArtifactBlob/#2\#L#3}{\texttt{\detokenize{#1}}}}

\hypersetup{
  hidelinks,
  colorlinks,
  breaklinks=true,
  urlcolor=color2,
  citecolor=color1,
  linkcolor=color1,
  bookmarksopen=false,
  pdftitle={Formal Foundations and Proof-Carrying Certificates for q-ary Covering Codes in Lean 4},
  pdfauthor={Andreas Florath}
}

\JournalInfo{Preprint, arXiv-v1, June 2026}
\Archive{}
\PaperTitle{Formal Foundations and Proof-Carrying Certificates for \texorpdfstring{$q$}{q}-ary Covering Codes in Lean 4}
\Authors{Andreas Florath\orcidlink{0009-0001-6471-7372}\textsuperscript{1}}
\affiliation{\textsuperscript{1}Deutsche Telekom AG, \textit{Andreas.Florath@telekom.de}}
\newcommand{\keywordname}{Keywords}

\Abstract{Covering codes in finite Hamming spaces ask for small sets of words whose Hamming balls cover the whole space.  This paper presents a \Lean{} 4 formalization of the elementary theory of \qary{} covering codes, centered on certificate predicates for upper bounds, lower bounds, and exact covering numbers \(K_q(n,r)\).  The formalization proves the \qary{} Hamming-ball volume formula, the sphere-covering lower bound, elementary exact cases, product and relation rules, and selected small exact certificates.  It also demonstrates an end-to-end workflow for checking explicit upper bounds transcribed from van Laarhoven et al.\ (1989).  The accompanying database is proof-carrying: stored bounds have traces that replay to \Lean{} proofs of the corresponding upper- or lower-bound predicates.  The contribution is not new record bounds or a reproduction of known tables, but a reusable, auditable foundation for machine-checked covering-code certificates.}

\begin{document}

\maketitle
\tableofcontents

\section{Introduction}

A simple way to meet covering codes is through a rook-placement problem.  Place
some rooks on a \(q\times q\) board, and call a square covered if it is occupied
or lies in the same row or column as an occupied square.  The question is how few
rooks are needed to cover the board.  This is already a covering-code problem:
the board is the set of pairs \(\{0,\ldots,q-1\}^2\), and one rook covers all
positions that differ from its own position in at most one coordinate.  Higher
dimensional versions replace the board by \(q^n\) grid points and ask for a
small set of centers whose Hamming balls cover the whole grid.

Another classical example is the football-pool
problem~\cite{hamalainen1995football,habsieger1996ternary}.  Suppose each of
\(n\) matches has three possible outcomes, viewed from the perspective of the
listed home team: win, draw, or loss.  A betting ticket is therefore a word in a
ternary alphabet.  The covering-code version of the pool problem asks how many
tickets are needed to guarantee a prize-level result: whatever the actual match
outcomes are, at least one submitted ticket should have at most one wrong
prediction, or equivalently at least \(n-1\) correct predictions.  This happens
exactly when every possible outcome vector differs from some ticket in at most
one coordinate.  In modern notation this asks for \(K_3(n,1)\), the minimum size
of a ternary radius-one covering code.

These examples illustrate the general problem.  In the Hamming space over a
finite alphabet \(A\), a word of length \(n\) is an element of \(A^n\), and the
Hamming distance \(\distH(x,y)\) counts the number of coordinates in which two
words differ~\cite{hamming1950error}.  We write \(\N\) for the natural numbers
including \(0\).  For \(r\in\N\), an \(r\)-covering code is a subset
\(C \subseteq A^n\) such that every word \(x \in A^n\) lies within
Hamming distance \(r\) of some codeword \(c \in C\).  Writing \(q:=|A|\), the
\qary{} covering number is defined by
\[
  K_q(n,r) = \min\{\, |C| : C \subseteq A^n \text{ and } C \text{ is an } r\text{-cover}\,\}
\]
and is the minimum size of such a covering code~\cite{cohen1997covering}.  Thus
\(K_q(n,r)\) asks for the smallest number of Hamming balls of radius \(r\)
needed to cover the whole space.

The definition is elementary, but the known values and bounds are built from a
mixture of ingredients: explicit constructions, counting arguments, recursive
relations between parameters, computer searches, and tables of best known
bounds.  Even for small parameters, an exact statement \(K_q(n,r)=k\) usually
has two independent parts.  One must exhibit a covering code with \(k\)
codewords, and one must prove that no smaller code can cover.  These two parts
often come from different sources.

This makes covering codes a natural test case for formalized mathematics.  The
objects are finite and concrete, so many proposed upper bounds can be checked by
computation.  At the same time, the correctness of a covering-number claim
depends on precise definitions, edge cases, and reusable mathematical lemmas.
For example, one has to say exactly what happens for the empty alphabet
\(q=0\), for length \(n=0\), and for radii \(r\ge n\).  Informal tables usually suppress
such conventions; a theorem prover cannot.

The aim of this paper is to turn elementary covering-code theory into
machine-checkable certificates.  A certificate here means a theorem-prover object
which can be checked independently of the program, search procedure, or table
entry that suggested it.  An explicit code proves an upper bound only after the
theorem prover checks that its Hamming balls really cover the space.  A lower
bound proves a different statement: every possible code below a given size
fails.  When the two bounds meet, they combine to an exact covering number.

The formalization is written in Lean 4~\cite{demoura2021lean4}.  It provides
finite Hamming spaces, Hamming balls, covering predicates, upper-bound and
lower-bound certificates, exactness certificates, the \qary{} ball-volume formula,
the sphere-covering lower bound, structural constructions, and a proof-carrying
database of covering-code bounds.  The database stores not only numerical
bounds, but also traces explaining how they were obtained.  Replaying such a
trace in Lean reconstructs a proof of the corresponding upper or lower bound.
To the author's knowledge, this is the first proof-assistant development whose
central objects are \qary{} covering numbers and composable covering-code
certificates; the main paper-level result is this auditable certificate
interface rather than any single numerical bound.

A guiding design choice is to model words as functions from a coordinate type to an alphabet type.  In Lean notation, the general word type is represented as a dependent function type,
\[
  \texttt{Word}\ \iota\ \beta := (i : \iota) \to \beta\ i,
\]
and the ordinary \qary{} Hamming space of length \(n\) is the specialization
\[
  \Fin\ n \to \Fin\ q.
\]
This representation makes the \qary{} case simple while preserving a more general interface for finite coordinate types and dependent alphabets.  Closed balls and covering predicates are then defined above this word layer.  Covering predicates come in both set-valued and finset-valued forms, because mathematical statements are often cleaner over sets while exact covering-number certificates and finite computations require finsets.

A second design choice concerns how to represent the covering number \(K_q(n,r)\) in the theorem prover.  Defining it directly as a function would require committing to a specific minimum construction in Lean, with its own proof obligations.  Such a definition would also be \emph{noncomputable} in Lean's sense: logically well-defined, but not directly evaluable by the proof kernel.  Making this choice prematurely would lock the formalization into one design decision before it is clear which is most convenient.

Instead, the formalization represents the assertion \(K_q(n,r)=k\) as a \emph{predicate} --- a logical statement that \(k\) satisfies the two conditions that characterise the minimum.  A statement \(\leanref{KSpec}{CoveringCodes/CoveringNumber.lean}{33}\ r\ k\) asserts: (1)~there exists a finite radius-\(r\) covering code of size \(k\) (an explicit upper-bound witness), and (2)~every finite radius-\(r\) covering code has size at least \(k\) (a lower-bound proof).  Together these two conditions are equivalent to \(K_q(n,r)=k\), but without ever defining \(K_q\) as a function.  The \qary{} specialization is written \(\leanref{QaryKSpec}{CoveringCodes/CoveringNumber.lean}{40}\ q\ n\ r\ k\).

This approach matches the natural structure of covering-code proofs --- an explicit construction for the upper bound and a counting argument for the lower bound --- and keeps the formalization modular: results from different sources (explicit constructions, sphere-covering bounds, or later computational certificates) can be assembled independently.  The two component predicates are also available separately: \(\leanref{KUpper}{CoveringCodes/CoveringNumber.lean}{50}\ r\ k\) expresses \(K_q(n,r)\le k\) and \(\leanref{KLower}{CoveringCodes/CoveringNumber.lean}{64}\ r\ k\) expresses \(K_q(n,r)\ge k\), with a lemma \(\leanref{KSpec.ofUpperLower}{CoveringCodes/CoveringNumber.lean}{89}\) recombining them into an exact certificate whenever both bounds agree on the same value \(k\).

\textbf{Contributions.}
This paper is a foundation paper rather than a record-setting paper.  Its
contributions are the following.
\begin{enumerate}[label=(\arabic*)]
  \item It gives a Lean-compatible framework for finite Hamming spaces and
  \qary{} covering codes, with words represented as functions and \qary{} words as
  \(\Fin\ n \to \Fin\ q\).  The framework includes set-valued and
  finset-valued covering predicates and treats boundary cases such as \(q=0\),
  \(n=0\), and \(r\ge n\) inside the formal statements.
  \item It develops certificate-style predicates for upper bounds
  (\leanref{KUpper}{CoveringCodes/CoveringNumber.lean}{50}), lower bounds
  (\leanref{KLower}{CoveringCodes/CoveringNumber.lean}{64}), and exact covering
  numbers (\leanref{KSpec}{CoveringCodes/CoveringNumber.lean}{33}), together
  with lemmas that project and recombine these certificates.
  \item It formalizes the \qary{} Hamming-ball volume formula
  \[
    V_q(n,r)=\sum_{i=0}^{\min(r,n)} \binom{n}{i}(q-1)^i
    \qquad(q\ge 1),
  \]
  using a Lean-total definition, and derives the classical sphere-covering lower
  bound.  It also proves elementary exact cases, including radius zero and the
  near-diameter value \(K_q(n,n-1)=q\) for \(1\le n\).
  \item It formalizes product and neighbor-relation rules for transporting
  covering-code bounds, including length, alphabet, block, concatenation, and
  structural explicit-code transformations.  One uniform construction gives
  \[
    K_q(3,1)\le \lfloor q/2\rfloor^2+\lceil q/2\rceil^2
    = \lceil q^2/2\rceil .
  \]
  \item It provides a trace-based proof-carrying database and demonstrates the
  artifact workflow on explicit published codes.  Stored bounds replay to
  \Lean{} proofs of \leanref{QaryKUpper}{CoveringCodes/CoveringNumber.lean}{54}
  or \leanref{QaryKLower}{CoveringCodes/CoveringNumber.lean}{68}; selected exact
  certificates include \(K_2(5,1)=7\), \(K_2(6,2)=4\),
  \(K_3(3,1)=5\), and \(K_4(3,1)=8\).  The van Laarhoven case study yields
  checked upper-bound certificates for \(K_3(6,1)\le 73\),
  \(K_3(7,1)\le 186\), and \(K_3(8,1)\le 486\).
\end{enumerate}

The scope is intentionally limited.  The paper does not claim new best bounds for \(K_q(n,r)\) and does not attempt to reproduce the full tables of known covering-code values from the literature.  The case study shows that committed explicit code data transcribed from a publication can be made into a Lean certificate, and the database section shows that registering such a certificate as a primitive source is straightforward.  The selected small exact certificates are included because they exercise the same upper/lower/exactness interfaces beyond the trivial structural families; their authority comes from the Lean proof terms, not from any external table entry used to identify candidate triples.  The finite covering-code problem has a direct set-cover formulation: choose binary variables for candidate codewords and require every word to be covered by at least one selected radius-\(r\) ball.  This formulation is the natural bridge to future SAT and integer-programming certificates; the switchable proof mode in the case study is a first step in that direction, and extending it to solver-generated certificates is left for future work.

\paragraph{Artifact availability.}
The Lean artifact is available from the source repository at the following fixed
revision:
\begin{center}
\footnotesize
\artifactrepo{}\\[-0.2ex]
\artifactcommit{}
\end{center}
All source-code links in this manuscript refer to that revision rather than to a
moving branch.  Artifact paths are relative to the repository root.  The build
and query instructions are recorded in
\leanfileat{README.md}{13}.  The ordinary proof mode checks finite certificates
through Lean's kernel reduction path; the optional native proof mode is used only
when explicitly enabled and has the trust boundary described in
Section~\ref{subsec:db-trust-boundary}.

\paragraph{Artifact audit workflow.}
For routine referee inspection, the recommended first artifact check is the
native proof-mode workflow described there.  It compiles the command-line query
interface and the generated-table dependencies, then exercises representative
database queries.
The proof-mode switch affects only expensive finite covering checks.  In native
mode, those checks additionally rely on Lean's native compiled evaluator; in
kernel mode, the same checks use ordinary \texttt{decide}.  Kernel mode is the
stronger replay path for those finite leaves, but the largest explicit-code files
require the resources shown in Table~\ref{tab:proof-mode-resources}.  The
remaining components (handwritten theory, generated trace typing, and trace
replay) are typechecked by Lean in the same way in both modes.

\section{Finite Hamming Spaces}\label{sec:hamming-spaces}

This section fixes the ambient objects used throughout the paper.  The mathematics is standard: the points of a finite Hamming space are finite words, and the distance between two words is the number of coordinates in which they differ.  The formalization is slightly more general than the usual \qary{} presentation, because many elementary lemmas do not depend on the alphabet being constant across coordinates.  The terminology and notation for finite Hamming spaces, Hamming distance, Hamming balls, shells, covering radius, and the covering number \(K_q(n,r)\) follow the standard conventions of coding and covering-code theory; for general coding-theory background see MacWilliams and Sloane~\cite{macwilliams1977theory} and van Lint~\cite{vanlint1999introduction}, and for covering-code notation see Cohen et al.~\cite[Sec.~2.1]{cohen1997covering}.

\subsection{Coordinate types and alphabets}

Let \(\iota\) be a finite type of coordinates.  For each coordinate \(i:\iota\), let \(\beta_i\) be the alphabet available at that coordinate.  A word over these coordinate alphabets is a dependent function
\[
  x : \prod_{i:\iota} \beta_i .
\]
In Lean this is represented directly by a dependent function type:
\begin{lstlisting}
abbrev (*@\leanref{Word}{CoveringCodes/Basic.lean}{20}@*) (ι : Type u) (β : ι → Type v) : Type _ :=
  (i : ι) → β i
\end{lstlisting}
This choice keeps the formal interface close to the mathematical notation.  A word \(x\) is applied to a coordinate \(i\), and the paper notation \(x_i\) corresponds to the Lean expression \texttt{x i}.

The assumptions needed for this layer are explicit.  The coordinate type must be finite, because Hamming distance is a finite count over coordinates.  Equality at every coordinate alphabet must be decidable, because the set of positions at which two words differ is defined by a decidable predicate.  When cardinalities of word spaces are used, the coordinate alphabets are also assumed finite.

Keeping the coordinate type abstract is useful later.  For example, the product construction for two codes is naturally stated over a sum of coordinate types before being transported to the usual \(\Fin(n_1+n_2)\) representation.  Thus the \qary{} case is the main specialization, not the only setting in which the elementary API is valid.

\subsection{The \qary{} specialization}

For \(q,n\in\N\), the \qary{} Hamming space of length \(n\) is
\[
  \mathcal H_q(n) := (\Fin\ q)^{\Fin\ n},
\]
which we also write informally as \((\Fin\ q)^n\).  In Lean this specialization is named as follows:
\begin{lstlisting}
abbrev (*@\leanref{QaryWord}{CoveringCodes/Basic.lean}{24}@*) (q n : ℕ) : Type :=
  Fin n → Fin q
\end{lstlisting}
Thus \texttt{QaryWord q n} is the type of length-\(n\) words over an alphabet with \(q\) symbols.

The definitions do not build in assumptions such as \(q>0\) or \(n>0\).  For example, \(\Fin 0\) is the empty type; hence there are no words of positive length over the alphabet \(\Fin 0\), while there is still one empty word when \(n=0\).  This agrees with the finite-type cardinality identity
\[
  |\Fin\ n \to \Fin\ q| = q^n,
\]
including the convention \(0^0=1\).  Later theorems add nonemptiness hypotheses, such as \(q\ge 1\), \(q\ge 2\), or \(n\ge 1\), exactly where they are mathematically needed.

Both \texttt{Word} and \texttt{QaryWord} are introduced with Lean's \texttt{abbrev} keyword rather than \texttt{def}, which makes them \emph{transparent} to the type-checking kernel.  As a result, \texttt{QaryWord q n} is definitionally equal to \texttt{Word (Fin n) (fun \_ : Fin n => Fin q)}, and every general theorem about \texttt{Word~\(\iota\)~\(\beta\)} applies to \qary{} words directly, without any explicit coercion or type cast.

This two-layer design --- a general layer over abstract \(\iota\) and \(\beta\),
and a \qary{} specialization with explicit \(q\), \(n\), \(r\), \(k\) --- is used
throughout the formalization.  General lemmas such as triangle inequalities,
ball monotonicity, radius monotonicity of covers, and the product construction
are proved once over arbitrary finite coordinate types.  Concrete covering-code
statements are then stated with \qary{} names such as
\texttt{QaryKSpec q n r k}, so the mathematically meaningful parameters remain
visible.  Because the \qary{} layer is transparent to Lean's kernel, this
specialization does not introduce a second proof layer.

\subsection{Hamming distance}\label{subsec:hamming-distance}

For two words \(x,y : \prod_{i:\iota}\beta_i\), the Hamming distance is
\[
  \distH(x,y)
  = \left|\{\, i:\iota : x_i \ne y_i \,\}\right|.
\]
The formalization reuses \mathlib{}'s Hamming-distance definition and exposes it
under the project namespace as \leanref{dist}{CoveringCodes/Basic.lean}{44}.
The codomain is \(\N\), not \(\mathbb R\).  This is the natural form for the elementary covering-code arguments: balls are defined by inequalities between natural numbers, exact-distance layers are finite sets, and all cardinality estimates remain in natural-number arithmetic.

The basic distance API consists of the following facts:
\begin{align*}
  \distH(x,x) &= 0, \\
  \distH(x,y) &= \distH(y,x), \\
  \distH(x,z) &\le \distH(x,y)+\distH(y,z), \\
  \distH(x,y) &\le |\iota|, \\
  \distH(x,y)=0 &\Longleftrightarrow x=y.
\end{align*}
For \qary{} words of length \(n\), the bound \(\distH(x,y)\le |\iota|\) specializes to \(\distH(x,y)\le n\).  In the formalization these facts are recorded by the expected self-distance, symmetry, triangle, cardinality, and zero-distance lemmas.  Together they provide the metric facts needed in the rest of the paper, while keeping the primary interface natural-number-valued.

\subsection{Balls and exact-distance layers}\label{subsec:balls-and-layers}

For a center word \(c\) and a radius \(r\in\N\), the closed Hamming ball of radius \(r\) around \(c\) is
\[
  B_r(c)=\{\,x : \distH(x,c)\le r\,\}.
\]
This is the geometric object from which covering codes are defined.  The Lean
definition \leanref{ball}{CoveringCodes/Balls.lean}{21} is the direct set of
words satisfying the distance inequality.
Consequently, membership in a ball is exactly the distance inequality:
\[
  x\in B_r(c) \quad\Longleftrightarrow\quad \distH(x,c)\le r.
\]
The elementary ball API contains the following facts:
\begin{enumerate}[label=(\roman*)]
  \item the center belongs to every ball around itself, \(c\in B_r(c)\);
  \item radius-zero balls are singletons, \(B_0(c)=\{c\}\);
  \item balls are monotone in the radius, so \(r\le s\) implies \(B_r(c)\subseteq B_s(c)\);
  \item if \(r\ge |\iota|\) (which equals \(n\) in the \qary{} case), then \(B_r(c)\) is the whole word space.
\end{enumerate}

It is also useful to refer informally to exact-distance layers.  For \(i\in\N\), write
\[
  S_i(c)=\{\,x : \distH(x,c)=i\,\}.
\]
Then
\[
  B_r(c)=\bigcup_{i=0}^{\min(r,n)} S_i(c)
\]
in the \qary{} length-\(n\) case.  This shell notation is used only to explain
the ball-volume formula below.

The notation \(B_r(c)\) and \(S_i(c)\) agrees with the usual Hamming ball and shell notation used in covering-code theory; see Cohen et al.~\cite[Sec.~2.1]{cohen1997covering}.

\subsection{Cardinalities and ball volume}\label{subsec:cardinalities-and-volume}

When all coordinate alphabets are finite, the dependent word space has cardinality
\[
  \left|\prod_{i:\iota}\beta_i\right|
  = \prod_{i:\iota} |\beta_i|.
\]
For the uniform \qary{} case this becomes
\[
  |\mathcal H_q(n)| = |\Fin\ n \to \Fin\ q| = q^n.
\]
This identity is used repeatedly: it gives the size of the full-space code, the cardinality of the search space, and the right-hand side of the sphere-covering bound.

For a fixed center in \(\mathcal H_q(n)\), the number of words at distance exactly \(i\) is
\[
  |S_i(c)|=\binom{n}{i}(q-1)^i.
\]
Indeed, one first chooses the \(i\) coordinates at which the word differs from the center and then chooses one of the \(q-1\) non-center symbols in each chosen coordinate.  Summing over exact-distance layers gives the \qary{} Hamming-ball volume
\[
  V_q(n,r)=\sum_{i=0}^{\min(r,n)} \binom{n}{i}(q-1)^i
  \qquad(q\ge 1).
\]
This classical formula is the standard volume formula for \qary{} Hamming balls; see Cohen et al.~\cite[Sec.~2.1]{cohen1997covering}.
The Lean function \leanref{qaryBallVolume}{CoveringCodes/Bounds/Balls.lean}{42}
implements this closed form as a total expression in \(q,n,r\) and uses
natural-number truncated subtraction, so \(q-1=0\) when \(q=0\).  For
\(q\ge 1\) the two agree.  The linked
\leansource{cardinality theorem}{CoveringCodes/Bounds/Balls.lean}{269}
states that the formula equals the size of an actual finite ball.
Equivalently, for every center \(c:\Fin\ n\to\Fin\ q\),
\[
  |B_r(c)|=V_q(n,r).
\]
This theorem is used later to derive the sphere-covering lower bound.  The theorem is stated for an actual center word, so edge cases such as \(q=0,n>0\) are handled by the type itself: in that case there is no center to quantify over.

The preceding definitions separate the finite Hamming-space layer from the covering-code layer introduced next.  The next section adds codes, covering predicates, and certificate-style exact covering-number statements on top of these definitions.

\section{Covering Codes and Covering Numbers}\label{sec:covering-codes}

Section~\ref{sec:hamming-spaces} fixed words, Hamming distance, and balls.  This
section adds the formal predicates used to state covering claims.  The main
predicate says that a chosen set of centers covers the ambient space at a given
radius.  The \qary{} covering number is then represented, in the current
formalization, by a certificate-style exactness predicate rather than by an
immediately chosen noncomputable minimization operator.

This distinction is intentional.  Mathematically, it is convenient to write \(K_q(n,r)\) for the minimum size of a radius-\(r\) cover of \((\Fin\ q)^n\).  In a theorem prover, however, many elementary results are naturally proved by giving an upper-bound witness and a lower-bound argument.  The formal predicate used here records precisely these two pieces of information.  Thus a statement such as \(K_q(n,r)=k\) is represented first as the existence of a finite cover of size \(k\), together with a proof that every finite cover has size at least \(k\).

\subsection{Covering codes as subsets of Hamming space}\label{subsec:codes-as-subsets}

Using the notation of Section~\ref{sec:hamming-spaces}, fix a finite word space
\[
  W = \prod_{i:\iota} \beta_i
\]
with Hamming distance \(\distH\).  For \(C\subseteq W\) and \(r\in\N\), the
formal covering predicate is
\[
  \operatorname{Covers}(C,r)
  \quad:\Longleftrightarrow\quad
  \forall x\in W,\ \exists c\in C,\ \distH(x,c)\le r.
\]
Thus \(C\) is a radius-\(r\) covering code exactly when
\(\operatorname{Covers}(C,r)\) holds.
Equivalently,
\[
  W = \bigcup_{c\in C} B_r(c),
\]
where \(B_r(c)\) is the closed Hamming ball of radius \(r\) centered at \(c\).  This union-of-balls view is the form used in the sphere-covering bound, while the existential view is usually the most convenient one for Lean proofs.

Specializing \(W\) to \(\mathcal H_q(n)\), the same condition becomes
\[
  \operatorname{Covers}_q(C,n,r)
  \quad:\Longleftrightarrow\quad
  \forall x:\mathcal H_q(n),\ \exists c\in C,\ \distH(x,c)\le r.
\]
This predicate is uniform in \(q\), \(n\), and \(r\).  No assumption such as \(q>0\) or \(n>0\) is built in.  This is useful because edge cases, for example the empty alphabet or the length-zero word space, can be handled by ordinary finite-type reasoning rather than by separate informal conventions.

\subsection{Set-valued and finset-valued predicates}\label{subsec:set-and-finset-covers}

The formalization exposes two covering predicates:
\leanref{Covers}{CoveringCodes/Covers.lean}{22} for set-valued codes and
\leanref{CoversFinset}{CoveringCodes/Covers.lean}{26} for finset-valued codes.
Both are direct translations of the mathematical statement above, using Lean's
membership notation and propositional connectives.

The set-valued predicate is useful for general inclusion arguments.  For example, if \(C\subseteq D\) and \(C\) covers at radius \(r\), then \(D\) also covers at radius \(r\).  The finset-valued predicate is useful whenever cardinalities are part of the statement, as in exact covering-number results and computational certificates.

The two predicates agree after coercing a finset to a set; this is the lemma
\leanref{coversFinset_iff_coe}{CoveringCodes/Covers.lean}{30}.  It lets the
formalization move freely between set-style and finset-style reasoning.  In
particular, theorems whose proofs are conceptually about containment can be
stated for sets, while theorems whose conclusions involve \(|C|\) can be stated
for finsets.

\subsection{Elementary interpretation as set cover}\label{subsec:set-cover-view}

For fixed \(q,n,r\), every possible center \(c\in\mathcal H_q(n)\) determines a subset of the ambient space, namely its ball \(B_r(c)\).  Finding a smallest radius-\(r\) covering code is therefore exactly the problem of selecting as few of these balls as possible so that their union is the whole space:
\[
  \mathcal H_q(n) = \bigcup_{c\in C} B_r(c).
\]
This is a finite instance of the standard set-cover formulation~\cite{cohen1997covering}: the universe is \(\mathcal H_q(n)\), and the candidate subsets are the balls \(\{B_r(c) : c\in\mathcal H_q(n)\}\).  Here we use this only as an elementary modeling viewpoint, not as a complexity-theoretic hardness claim.  The set-cover viewpoint is not needed to define covering codes, but it clarifies two later parts of the paper.  First, the sphere-covering lower bound arises by comparing the cardinality of the full space with the total size of the balls selected by a code.  Second, computational upper-bound certificates are simply explicit choices of centers whose associated balls cover all words.

This paper uses the set-cover perspective only at the elementary level.  More sophisticated solver-based certificates, symmetry reductions, and integer-programming or SAT encodings are deliberately left for future work; Section~\ref{sec:computation} records the basic computational interface.

\subsection{The \qary{} covering number}\label{subsec:qary-covering-number}

The introduction used the standard notation \(K_q(n,r)\) for the least size of a
radius-\(r\) covering code in \(\mathcal H_q(n)\):
\[
  K_q(n,r)
  =
  \min\{\, |C| : C\subseteq \mathcal H_q(n),\ C\text{ is an }r\text{-cover}\,\}.
\]
This is standard notation~\cite[Sec.~2.1]{cohen1997covering}.
The Lean development does not make this minimum function the primary formal
object.  Its certificate interface uses linked predicates for
\leansource{upper}{CoveringCodes/CoveringNumber.lean}{54} and
\leansource{lower}{CoveringCodes/CoveringNumber.lean}{68} bounds, together with
\leansource{exactness}{CoveringCodes/CoveringNumber.lean}{40}.

The full-space construction gives the immediate upper bound
\[
  K_q(n,r) \le |\mathcal H_q(n)| = q^n.
\]
If \(r\ge n\), then a single word covers the entire \qary{} Hamming space, because every two length-\(n\) words have distance at most \(n\).  Under the nonemptiness assumption \(\mathcal H_q(n)\ne\varnothing\), this gives
\[
  K_q(n,r)=1
  \qquad (r\ge n).
\]
These are elementary examples of the upper-bound/lower-bound pattern used throughout the formalization.

\subsection{Certificate-style exactness}\label{subsec:kspec}

Instead of making a noncomputable definition of \(K_q(n,r)\) the central formal object, the current formalization uses an exactness predicate.  For a fixed ambient word type, the predicate
\[
  \operatorname{KSpec}(r,k)
\]
means that \(k\) is the exact minimum size of a radius-\(r\) cover.  It is defined by the conjunction of an upper-bound certificate and a lower-bound theorem:
\[
\begin{aligned}
  \operatorname{KSpec}(r,k)\quad :\Longleftrightarrow\quad
  &\exists C,\ |C|=k,\ C\text{ is an }r\text{-cover},\\
  &\text{and for every finite }D,\\
  &\qquad D\text{ is an }r\text{-cover}\Rightarrow k\le |D|.
\end{aligned}
\]
The corresponding Lean definition is:
\begin{lstlisting}
def (*@\leanref{KSpec}{CoveringCodes/CoveringNumber.lean}{33}@*) (r k : ℕ) : Prop :=
  ∃ C : Finset (Word ι β),
    C.card = k ∧
    CoversFinset C r ∧
    ∀ D : Finset (Word ι β), CoversFinset D r → k ≤ D.card
\end{lstlisting}
The \qary{} specialization
\leanref{QaryKSpec}{CoveringCodes/CoveringNumber.lean}{40} makes all four
numerical parameters explicit.  Thus a Lean theorem of type
\texttt{QaryKSpec q n r k} corresponds to the mathematical assertion
\(K_q(n,r)=k\).

This representation has two practical advantages.  First, it matches the usual structure of covering-code proofs: an explicit code proves \(K_q(n,r)\le k\), and a lower-bound argument proves \(K_q(n,r)\ge k\).  Second, it avoids forcing early design decisions about a global covering-number function, for example whether to define it by \texttt{Nat.find}, by a finite minimum over a finset of candidate cardinalities, or by an order-theoretic infimum.  Such a function can be added later and proved equivalent to \leanref{KSpec}{CoveringCodes/CoveringNumber.lean}{33}; the present formalization only needs the certificate form.

The two conjuncts of \leanref{KSpec}{CoveringCodes/CoveringNumber.lean}{33} are also exposed as standalone predicates.  The predicate \(\operatorname{KUpper}(r,k)\) expresses the upper bound \(K_q(n,r)\le k\): there exists a finite radius-\(r\) covering code of size at most \(k\).  The predicate \(\operatorname{KLower}(r,k)\) expresses the lower bound \(K_q(n,r)\ge k\): every finite radius-\(r\) covering code has size at least \(k\).  Their Lean definitions are:
\begin{lstlisting}
def (*@\leanref{KUpper}{CoveringCodes/CoveringNumber.lean}{50}@*) (r k : ℕ) : Prop :=
  ∃ C : Finset (Word ι β), C.card ≤ k ∧ CoversFinset C r

def (*@\leanref{KLower}{CoveringCodes/CoveringNumber.lean}{64}@*) (r k : ℕ) : Prop :=
  ∀ C : Finset (Word ι β), CoversFinset C r → k ≤ C.card
\end{lstlisting}
The three predicates are connected by linked
\leansource{projection}{CoveringCodes/CoveringNumber.lean}{72},
\leansource{combination}{CoveringCodes/CoveringNumber.lean}{89}, and
\leansource{uniqueness}{CoveringCodes/CoveringNumber.lean}{95} lemmas.  The
combination lemma is the key result: an upper-bound witness satisfies
\(|C|\le k\), while the lower-bound theorem applied to the same code forces
\(k\le|C|\), so \(\lvert C\rvert=k\) follows by antisymmetry.  The uniqueness
lemma confirms that the certificate predicate behaves like a numerical
covering-number value: two exact certificates for the same radius must agree on
their value.  The upper- and lower-bound predicates also satisfy linked
\leansource{monotonicity}{CoveringCodes/CoveringNumber.lean}{110} and
\leansource{antitonicity}{CoveringCodes/CoveringNumber.lean}{116} lemmas.  Their
\qary{} specializations make all four parameters explicit, in the same style as
\texttt{QaryKSpec}.

\subsection{Examples of exactness statements}\label{subsec:exactness-examples}

The elementary exact values proved later in the paper are naturally expressed using \texttt{KSpec} and \texttt{QaryKSpec}.  For example, the zero-radius theorem says that every word must be selected as a codeword:
\[
  K_q(n,0)=q^n.
\]
In certificate form, the linked
\leansource{general theorem}{CoveringCodes/SmallCases/ZeroRadius.lean}{41}
states that the whole finite word space is the unique minimum-size radius-zero
cover.
The upper-bound witness is the full word space, and the lower-bound proof shows that a radius-zero cover must contain every word.

Another example is the near-diameter theorem:
\[
  K_q(n,n-1)=q
  \qquad (n\ge 1).
\]
The corresponding linked
\leansource{Lean theorem}{CoveringCodes/SmallCases/RadiusNMinusOne.lean}{202}
has the expected
\leanref{QaryKSpec}{CoveringCodes/CoveringNumber.lean}{40} conclusion.
The upper-bound witness is the code of constant \qary{} words, while the lower bound shows that every radius-\((n-1)\) cover must contain at least \(q\) words.  This theorem illustrates why the certificate-style API is convenient: the proof object contains both the construction and the optimality argument.

\subsection{Specializations and naming discipline}\label{subsec:specializations}

Some classical problems are obtained by fixing the alphabet size and radius.  For example, the ternary radius-one problem is the specialization \(K_3(n,1)\).  It may be useful to introduce an abbreviation for such a specialization in examples or in a later module, but the core formalization does not depend on doing so.  The formal library is organized around arbitrary \(q\), \(n\), and \(r\), and the \qary{} specialization itself is only a specialization of the more general finite-word interface.

This naming discipline matters for the scope of the paper.  The first goal is not to formalize one named application, but to build a reusable foundation for finite Hamming-space covering codes.  Application-specific terminology can then be layered on top without changing the underlying definitions.

The next section records elementary consequences of these definitions: monotonicity in the radius, full-space and singleton covers, zero-radius exactness, and near-diameter exactness.  These results are mathematically simple, but they are important tests that the formal interface has the right shape.

\section{Elementary General Theorems}\label{sec:elementary-theorems}

The previous two sections introduced the finite Hamming-space interface and the covering-code predicates.  This section records the first general theorems built on top of that interface.  None of the results is intended to be mathematically surprising.  Their role is foundational: they test that the definitions have the right shape, establish the standard edge cases, and provide small reusable lemmas for later bounds and constructions.

There are two kinds of statements in this section.  The first kind concerns the behavior of balls and covers under changes of radius.  These are structural lemmas used throughout the formalization.  The second kind packages elementary exact values of covering numbers in certificate form.  These exact values are simple, but they are important because they exercise both halves of the exactness predicate introduced in Section~\ref{subsec:kspec}: an explicit covering code and a proof that no smaller code can cover.

\subsection{Ball lemmas}\label{subsec:elementary-ball-lemmas}

The following lemmas are immediate consequences of the ball definition from Section~\ref{subsec:balls-and-layers} and the basic Hamming-distance API.

Ball membership is just the defining distance inequality:
\[
  x\in B_r(c) \quad\Longleftrightarrow\quad \distH(x,c)\le r.
\]
The Lean simp lemma is
\leanref{mem_ball_iff}{CoveringCodes/Balls.lean}{28}.  The same file proves
that the center is contained in every ball around itself
(\leanref{center_mem_ball}{CoveringCodes/Balls.lean}{39}),
that radius-zero balls are singletons
(\leanref{ball_zero}{CoveringCodes/Balls.lean}{33}), and that balls are monotone
in the radius (\leanref{ball_subset_ball_of_le}{CoveringCodes/Balls.lean}{44}):
\[
  c\in B_r(c), \qquad B_0(c)=\{c\}, \qquad
  r\le s \Rightarrow B_r(c)\subseteq B_s(c).
\]
These lemmas are small, but they are the local facts that make later covering
proofs readable.  For example, radius monotonicity of covers is simply ball
monotonicity expressed in the existential form of the covering predicate.

\subsection{Radius monotonicity of covers}\label{subsec:cover-radius-monotonicity}

If a code covers at radius \(r\), then it covers at every larger radius.  For a set-valued code \(C\subseteq W\), the statement is
\[
  r\le s \quad\text{and}\quad \operatorname{Covers}(C,r)
  \quad\Longrightarrow\quad
  \operatorname{Covers}(C,s).
\]

The formalization provides both a linked
\leansource{set-valued theorem}{CoveringCodes/Covers.lean}{35} and a linked
\leansource{finset-valued theorem}{CoveringCodes/Covers.lean}{43}.  The
finset-valued theorem is obtained by transporting the set-valued theorem through
the equivalence between the two covering predicates after coercing a finset to a
set.

Mathematically, this lemma is the source of the usual antitonicity of covering numbers in the radius:
\[
  r\le s \quad\Longrightarrow\quad K_q(n,s)\le K_q(n,r).
\]
In the certificate-style API used here, the theorem is most directly used as an upper-bound transfer: a radius-\(r\) covering certificate immediately gives a radius-\(s\) covering certificate of the same size.  A global noncomputable covering-number function can later expose the displayed inequality as a named theorem.

A closely related monotonicity principle concerns the code itself rather than the radius: adding codewords cannot destroy the covering property.  If \(C\subseteq D\) and \(C\) is an \(r\)-cover, then \(D\) is also an \(r\)-cover.  This lemma is not mathematically difficult, but it is a useful companion to radius monotonicity because many constructions enlarge or transform existing codes.

\subsection{Full-space covers}\label{subsec:full-space-covers}

The simplest covering code is the whole ambient space.  Let \(W\) be any finite word space.  The full-space code covers at every radius:
\[
  \operatorname{CoversFinset}(W,r).
\]

The linked \leansource{Lean theorem}{CoveringCodes/Covers.lean}{54} is stated
using the universal finset.
In the \qary{} case this gives the elementary upper bound
\[
  K_q(n,r)\le |\mathcal H_q(n)|=q^n.
\]
This upper bound is usually far from sharp, but it has two important formal uses.  First, it shows that the family of finite covering codes is nonempty whenever one wants to define \(K_q(n,r)\) as a minimum.  Second, it provides the upper-bound half of the radius-zero exactness theorem.

\subsection{Large-radius covers}\label{subsec:large-radius-covers}

The next elementary observation is that the maximum possible Hamming distance between two words is the number of coordinates.  If \(|\iota|\) denotes the cardinality of the coordinate type, then
\[
  \distH(x,y)\le |\iota|
\]
for all \(x,y\in W\).  Consequently, a single center covers the whole word space at radius \(|\iota|\):
\[
  \{c\}\text{ is a }|\iota|\text{-cover}.
\]
The linked \leansource{Lean theorem}{CoveringCodes/SmallCases/LargeRadius.lean}{33}
formalizes this singleton cover.
By radius monotonicity, the same singleton covers at every larger radius
(\leansource{large-radius version}{CoveringCodes/SmallCases/LargeRadius.lean}{57}).
There is also a
\leansource{version for arbitrary nonempty codes}{CoveringCodes/SmallCases/LargeRadius.lean}{45};
its proof chooses one element \(c\in C\) and ignores all other codewords.

Specialized to \(\mathcal H_q(n)\), the coordinate type is \(\Fin n\), so \(|\iota|=n\).  Hence, whenever the ambient space is nonempty and \(r\ge n\), one codeword is enough:
\[
  K_q(n,r)=1
  \qquad\text{if }\mathcal H_q(n)\ne\varnothing\text{ and }r\ge n.
\]
The nonemptiness hypothesis is essential in a fully formal statement.  If the ambient space is empty, then the empty code already covers vacuously, so the exact minimum is \(0\), not \(1\).  For \qary{} spaces, \(\mathcal H_q(n)\) is nonempty exactly when \(q>0\) or \(n=0\).

The same observation can also be stated geometrically: for any center \(c\) and any radius \(r\ge |\iota|\), the ball \(B_r(c)\) is the whole word space.  The formalization currently uses the equivalent covering-code formulation because it is the form needed for exactness certificates.

\subsection{Zero-radius exactness}\label{subsec:zero-radius-exactness}

At radius zero, a codeword covers only itself.  Therefore a radius-zero covering code must contain every word of the ambient space.  In the \qary{} case this gives the exact value
\[
  K_q(n,0)=q^n.
\]

The corresponding classical \qary{} theory records the trivial values \(K_q(n,0)=q^n\), \(K_q(n,n)=1\), and the near-diameter value \(K_q(n,n-1)=q\) under the usual nonempty-alphabet convention; see Cohen et al.~\cite[Sec.~3.7]{cohen1997covering}.
The Lean interface uses natural-number parameters throughout, so the boundary
cases are part of the formal statements rather than informal side conditions.
Table~\ref{tab:edge-cases} summarizes the conventions and the corresponding
Lean names.

\begin{table*}[t]
\centering
\footnotesize
\setlength{\tabcolsep}{3pt}
\begin{tabular}{@{}p{0.15\textwidth}p{0.48\textwidth}p{0.31\textwidth}@{}}
\toprule
Case & Formal behavior & Lean declarations \\
\midrule
All \(q,n,r\in\mathbb N\) &
No positivity assumption is built into the \qary{} predicates; the word space is
the finite type \(\Fin n\to\Fin q\). &
\texttt{QaryWord}, \texttt{QaryKSpec}, \texttt{QaryKUpper},
\texttt{QaryKLower} \\
\(q=0\) &
The word space is nonempty exactly when \(n=0\).  For \(n>0\), covering
conditions over the empty space are vacuous, and lower bound \(0\) is the
database fallback. &
\texttt{qaryWord\_nonempty\_iff}, \texttt{zeroLower\_valid},
\texttt{trivialUpper\_valid} \\
\(r=0\) &
Exact for all \(q,n\): \(K_q(n,0)=q^n\).  This includes the empty-space value
\(K_0(n,0)=0\) for \(n>0\) and \(K_q(0,0)=1\). &
\texttt{covers\_zero\_iff\_univ\_subset},
\texttt{qaryKSpec\_zero\_radius} \\
\(n=0\) &
There is one empty word, so the exact value is \(K_q(0,r)=1\) for every \(r\).
This follows from zero-radius exactness at \(r=0\) and from the large-radius
source bounds for \(r\ge0\). &
\texttt{qaryWord\_nonempty\_iff}, \texttt{qaryKSpec\_zero\_radius},
\texttt{largeRadiusLower\_valid}, \texttt{largeRadiusUpper\_valid} \\
\(r\ge n\) &
A singleton covers whenever the word space has a center, giving exact value
\(1\) for \(n=0\) or \(q>0\).  For \(q=0,n>0\), the space is empty and the
trivial upper/lower bounds give value \(0\). &
\texttt{singleton\_covers\_of\_card\_le},
\texttt{largeRadiusLower\_valid}, \texttt{largeRadiusUpper\_valid} \\
\(r=n-1\) &
Exact for all \(q\) when \(1\le n\): \(K_q(n,n-1)=q\).  The \(n=0\) case is
not part of this theorem and is handled by the radius-zero case. &
\texttt{qaryKSpec\_radius\_nMinusOne},
\texttt{radiusNMinusOneLower\_valid},
\texttt{radiusNMinusOneUpper\_valid} \\
\bottomrule
\end{tabular}
\caption{Boundary cases included by the all-natural-number \qary{} interface.}
\label{tab:edge-cases}
\end{table*}

The linked
\leansource{key formal lemma}{CoveringCodes/SmallCases/ZeroRadius.lean}{24}
characterizes radius-zero covers by containment of the universal finset.
This provides the lower-bound half of the exactness certificate.

The linked
\leansource{exactness theorem}{CoveringCodes/SmallCases/ZeroRadius.lean}{41} is
packaged directly in certificate form.
The upper-bound witness is \texttt{Finset.univ}, the finset of all words.  The lower-bound part uses the zero-radius characterization above: if \(D\) is any radius-zero cover, then it contains every word, and therefore
\[
  |W|\le |D|.
\]

This theorem is a useful example of the intended proof pattern for exact covering numbers.  The construction and the lower bound are separate arguments, but the \texttt{KSpec} theorem combines them into a single machine-checkable certificate of exactness.

\subsection{Length-zero and length-one consequences}\label{subsec:length-zero-one}

Several edge cases follow immediately from the large-radius and zero-radius results.  They are worth spelling out because they prevent hidden informal assumptions about nonempty alphabets or positive word lengths.

For length zero, the \qary{} Hamming space contains exactly one word: the empty function from the empty coordinate type to the alphabet.  This remains true for \(q=0\), because there is a unique function from the empty type to \(\Fin 0\).  Thus
\[
  |\mathcal H_q(0)|=q^0=1
\]
for every \(q\), including \(q=0\).  Since the only word is at distance zero from itself, one codeword covers for every radius:
\[
  K_q(0,r)=1.
\]
The nonemptiness requirement in the large-radius singleton theorem is therefore satisfied at \(n=0\), so this agrees with both zero-radius exactness and the large-radius result.

For length one, radius zero requires all symbols:
\[
  K_q(1,0)=q.
\]
For positive radius, one selected symbol covers every one-letter word, provided the \qary{} space is nonempty:
\[
  K_q(1,r)=1
  \qquad(q\ge 1,\ r\ge 1).
\]
If \(q=0\) and \(n=1\), the ambient word space is empty and the exact covering number is \(0\).  These degenerate cases are not exceptions to the formal definitions; they are ordinary consequences of representing \qary{} words as functions \(\Fin n\to\Fin q\).

\subsection{The near-diameter value}\label{subsec:near-diameter-value}

The most substantial elementary exact value in this initial formalization is the
\leansource{near-diameter case}{CoveringCodes/SmallCases/RadiusNMinusOne.lean}{202}
\[
  K_q(n,n-1)=q
  \qquad(n\ge 1).
\]
Common informal presentations often state this for \(q\ge 2\), but the formal
theorem does not need that hypothesis.  For \(q=0\) and \(n\ge 1\), the ambient
space is empty and the value is \(0=q\); for \(q=1\), the ambient space has one
word and the value is \(1=q\).

The upper bound uses the finite set of constant words.  Its linked
\leansource{cardinality}{CoveringCodes/SmallCases/RadiusNMinusOne.lean}{76} and
\leansource{covering}{CoveringCodes/SmallCases/RadiusNMinusOne.lean}{117} lemmas
establish the construction.  The lower bound is the linked
\leansource{missing-symbol theorem}{CoveringCodes/SmallCases/RadiusNMinusOne.lean}{158}:
any radius-\((n-1)\) cover has at least \(q\) codewords.  Combining these two
parts gives the exact linked
\leansource{certificate}{CoveringCodes/SmallCases/RadiusNMinusOne.lean}{202}.
This theorem is a useful first nontrivial test of the linked
\leanref{QaryKSpec}{CoveringCodes/CoveringNumber.lean}{40} interface.

These elementary results are deliberately modest.  Their importance is that they establish the basic proof pattern used throughout the rest of the paper: define a finite code, prove that it covers, and separately prove that any covering code must have at least a certain size.  The next section applies the same pattern at a larger scale by counting the union of Hamming balls around an arbitrary covering code.

\section{Hamming Balls and the Sphere-Covering Bound}\label{sec:sphere-bound}

This section contains the two main analytical results of the formalization.  The first is a proof of the closed-form ball volume formula introduced in Section~\ref{subsec:cardinalities-and-volume}, which is needed as a lemma for everything that follows.  The second is the classical sphere-covering lower bound, which asserts that the number of codewords in any covering code is at least \(\lceil q^n / V_q(n,r) \rceil\).  Together these two results form the analytical core of the library and the primary input to the certified lower-bound database described in Section~\ref{sec:computation}.

\subsection{Proof of the ball volume formula}\label{subsec:ball-volume-proof}

Section~\ref{subsec:cardinalities-and-volume} introduced the \qary{} Hamming-ball volume
\[
  V_q(n,r)=\sum_{i=0}^{\min(r,n)}\binom{n}{i}(q-1)^i .
\]
The linked theorem
\leanref{qaryBallVolume_eq_card}{CoveringCodes/Bounds/Balls.lean}{269}
formalizes the classical formula
\(|B_r(c)|=V_q(n,r)\)~\cite[Sec.~2.1]{cohen1997covering} by double
induction on \(n\) and \(r\).  The base cases \(n=0\) and \(r=0\) are
immediate.

The inductive step is the usual first-coordinate split.  A word in
\(B_{r+1}(c)\) either agrees with \(c\) in the first coordinate and leaves a
tail in \(B_{r+1}(c_{\mathrm{tail}})\), or differs there and chooses one of the
\(q-1\) non-center symbols while leaving a tail in \(B_r(c_{\mathrm{tail}})\).
Thus the ball cardinalities satisfy
\[
  |B_{r+1}(c)| =
  |B_{r+1}(c_{\mathrm{tail}})| +
  (q-1)\,|B_r(c_{\mathrm{tail}})|.
\]
The corresponding arithmetic recurrence is:
\begin{lstlisting}
lemma (*@\leanref{qaryBallVolume_succ_succ}{CoveringCodes/Bounds/Balls.lean}{179}@*) (q n r : ℕ) :
    qaryBallVolume q (n + 1) (r + 1) =
    qaryBallVolume q n (r + 1) + (q - 1) * qaryBallVolume q n r
\end{lstlisting}
The main formal nuisance is that the two sums naturally have different upper
bounds because of \(\min(r,n)\).  The Lean proof rewrites the volume formula to
an equivalent sum over \(0,\ldots,r+1\), using \(\binom{n}{i}=0\) for \(i>n\);
after that, Pascal's identity aligns the summands and the induction hypothesis
matches the ball recurrence.  Positivity of
\leanref{qaryBallVolume}{CoveringCodes/Bounds/Balls.lean}{42} then follows from
the \(i=0\) summand and is used in the ceiling-division form of the
sphere-covering bound below.

\subsection{The sphere-covering lower bound}\label{subsec:sphere-bound}

The sphere-covering lower bound is the standard counting argument for covering codes: each codeword covers at most \(V_q(n,r)\) words, so a radius-\(r\) covering code must contain enough codewords to cover all \(q^n\) words \cite[Sec.~2.1]{cohen1997covering}.  The formal statement avoids natural-number division: the bound is phrased as a product inequality.

\textbf{Intuition.}  Let \(C\subseteq\mathcal H_q(n)\) be a finite covering code of radius \(r\).  Each codeword \(c\in C\) is associated with a ball \(B_r(c)\), and the balls together cover all \(q^n\) words.  Even if the balls were pairwise disjoint --- the most favorable case for the code --- they could account for at most \(|C|\cdot V_q(n,r)\) words in total.  Since they must account for all \(q^n\) words:
\[
  q^n \le |C|\cdot V_q(n,r).
\]
In practice the balls overlap, which means a given word may be covered multiple times.  The inequality is therefore conservative: it gives a lower bound, not an equality.

\textbf{Formal statement.}  The theorem is phrased as an inequality between \(q^n\) and the product \(|C|\cdot V_q(n,r)\):
\begin{lstlisting}
theorem (*@\leanref{sphere_covering_bound}{CoveringCodes/Bounds/SphereCoveringBound.lean}{40}@*) {q n r : ℕ}
    (C : Finset (QaryWord q n)) (hC : CoversFinset C r) :
    q ^ n ≤ C.card * qaryBallVolume q n r
\end{lstlisting}

\textbf{Formal proof.}  The proof follows the intuitive argument as a single \texttt{calc} chain over finset cardinalities:
\begin{lstlisting}
calc q ^ n
    = Finset.univ.card                            := ...
  _ ≤ (C.biUnion (fun c => ballFinset c r)).card := Finset.card_le_card hcover
  _ ≤ Finset.sum C (fun c => (ballFinset c r).card)     := Finset.card_biUnion_le
  _ = Finset.sum C (fun c => qaryBallVolume q n r)       := Finset.sum_congr rfl (...)
  _ = C.card * qaryBallVolume q n r              := by rw [Finset.sum_const, smul_eq_mul]
\end{lstlisting}

\subsection{Lower bound via ceiling division}\label{subsec:ceiling-division}

The product inequality \(q^n\le |C|\cdot V_q(n,r)\) can be rearranged to \(|C|\ge\lceil q^n / V_q(n,r)\rceil\), giving an explicit computable lower bound.  The ceiling-division version is the natural integer-arithmetic form of the same sphere-covering bound.  In natural-number arithmetic, ceiling division is defined as follows:
\[
  \left\lceil \frac{a}{b} \right\rceil = \frac{a+b-1}{b}
  \qquad (b>0),
\]
where the division on the right is natural-number floor division.  The Lean definition is:
\begin{lstlisting}
def (*@\leanref{natCeilDiv}{CoveringCodes/Database/Arithmetic.lean}{10}@*) (a b : ℕ) : ℕ :=
  (a + b - 1) / b
\end{lstlisting}
The convention \texttt{natCeilDiv~a~0~=~0} is adopted for the zero-denominator case, which never arises in the sphere-covering application because \(V_q(n,r)\ge 1\) always holds.

The key arithmetic lemma relating ceiling division to product inequalities is:
\begin{lstlisting}
theorem (*@\leanref{natCeilDiv_le_of_le_mul}{CoveringCodes/Database/Arithmetic.lean}{14}@*) {a b c : ℕ}
    (hb : 0 < b) (h : a ≤ c * b) :
    natCeilDiv a b ≤ c
\end{lstlisting}
This says that if \(a\le c\cdot b\) and \(b>0\), then
\(\lceil a/b\rceil\le c\).  The proof is a short calculation in natural-number
arithmetic, using the standard characterization of strict inequalities for
natural-number division and the fact that \(\lceil a/b\rceil-1 < a/b\).

The sphere-covering lower bound is encapsulated as a computable function:
\begin{lstlisting}
def (*@\leanref{sphereLower}{CoveringCodes/Database/Sources/SphereCovering.lean}{12}@*) (q n r : ℕ) : ℕ :=
  natCeilDiv (q ^ n) (qaryBallVolume q n r)
\end{lstlisting}
Its validity as a lower-bound certificate is established by:
\begin{lstlisting}
theorem (*@\leanref{sphereLower_valid}{CoveringCodes/Database/Sources/SphereCovering.lean}{16}@*) (q n r : ℕ) :
    QaryKLower q n r (sphereLower q n r)
\end{lstlisting}
The proof applies the natural-number ceiling-division lemma with \(a:=q^n\),
\(b:=V_q(n,r)\), and \(c:=|C|\) for an arbitrary covering code \(C\).
Positivity of the denominator comes from the ball-volume positivity lemma, and
the product inequality comes from the sphere-covering bound.  Thus
\texttt{sphereLower\_valid} is a direct consequence of the two main results of
this section.

The results in this section are mathematically standard.  Their value in the formalization is that they provide a machine-checked bridge between the explicit ball-volume formula and the certificate-style lower-bound predicates used by the database.  The next section applies the same pattern at the construction level by combining covering codes via a product operation.

\section{Elementary Constructions}\label{sec:constructions}

This section presents the product construction for covering codes.  The idea is classical: given two codes, one covering a left coordinate block and one covering a right coordinate block, their Cartesian product covers the combined block at the sum of the two radii.  This is the formal version of the standard direct-sum construction for covering codes, in which covering radii add under products; see Cohen et al.~\cite[Thm.~3.2.1]{cohen1997covering}.  The formalization states and proves this in the general setting of abstract coordinate types, with no restriction to uniform alphabets or \qary{} codes.  The resulting theorem is then a direct building block for the database and for future constructions that involve lengthened or concatenated codes.

The section follows the structure established in earlier sections.  First, the notion of a product word over a sum of coordinate types is introduced.  Second, the additivity of Hamming distance across the two coordinate blocks is established.  Third, the product code is defined and the main covering theorem is proved.

\subsection{Sum-type words}\label{subsec:sum-words}

Let \(\iota\) and \(\kappa\) be finite coordinate types, with alphabet families \(\beta_1 : \iota \to \mathsf{Type}\) and \(\beta_2 : \kappa \to \mathsf{Type}\).  A word over the disjoint union \(\iota \oplus \kappa\) assigns a symbol to each coordinate in \(\iota\) and to each coordinate in \(\kappa\).  Here and throughout this section, \(\oplus\) denotes the type-level disjoint union of coordinate types, Lean's \texttt{Sum}, not XOR or a vector-space/group operation.  Given a word \(x_1\) over \(\iota\) and a word \(x_2\) over \(\kappa\), their concatenation is the word
\[
  \operatorname{sumWord}(x_1,x_2) : (s : \iota \oplus \kappa) \to (\beta_1 \oplus \beta_2)(s)
\]
defined by
\begin{align*}
  \operatorname{sumWord}(x_1,x_2)(\mathsf{inl}\;i) &= x_1(i), \\
  \operatorname{sumWord}(x_1,x_2)(\mathsf{inr}\;j) &= x_2(j).
\end{align*}
Here \(\mathsf{inl}\) and \(\mathsf{inr}\) are the two constructors of the sum type \(\iota\oplus\kappa\), and the combined alphabet at position \(s\) is the family \(\mathtt{Sum.elim}\ \beta_1\ \beta_2\ s\) which returns \(\beta_1 i\) when \(s = \mathsf{inl}\;i\) and \(\beta_2 j\) when \(s = \mathsf{inr}\;j\).

The Lean definition mirrors this case analysis directly:
\begin{lstlisting}
def (*@\leanref{sumWord}{CoveringCodes/Constructions/Product.lean}{42}@*) (x1 : (i : iota) → beta1 i) (x2 : (j : kappa) → beta2 j) :
    (s : Sum iota kappa) → Sum.elim beta1 beta2 s :=
  fun s =>
    match s with
    | Sum.inl i => x1 i
    | Sum.inr j => x2 j
\end{lstlisting}
Conversely, every sum word decomposes into its left projection
\(x_1(i) := x(\mathsf{inl}\;i)\) and its right projection
\(x_2(j) := x(\mathsf{inr}\;j)\), and then reassembles as
\(x = \operatorname{sumWord}(x_1, x_2)\).  This reassembly identity is proved by
pointwise case analysis on the sum type and plays a central role in the main
covering theorem.

\subsection{Additivity of Hamming distance}\label{subsec:dist-additivity}

The key structural property of the product construction is that Hamming distance decomposes additively along the left and right coordinate blocks.  Formally, for any two sum words,
\[
\begin{aligned}
  &\distH\!\left(\operatorname{sumWord}(x_1,x_2),
                 \operatorname{sumWord}(y_1,y_2)\right)\\
  &\qquad= \distH(x_1,y_1) + \distH(x_2,y_2).
\end{aligned}
\]
This is the content of the lemma
\begin{lstlisting}
lemma (*@\leanref{hammingDist_sumWord}{CoveringCodes/Constructions/Product.lean}{50}@*)
    (x1 y1 : Word iota beta1) (x2 y2 : Word kappa beta2) :
    hammingDist (sumWord x1 x2) (sumWord y1 y2) =
    hammingDist x1 y1 + hammingDist x2 y2
\end{lstlisting}

The proof decomposes the filter over \(\iota\oplus\kappa\) that defines Hamming distance into two disjoint parts.  Write
\(\Delta(u,v)=\{\,t \mid u(t)\ne v(t)\,\}\), and let \(\mathsf{inl}[-]\) and
\(\mathsf{inr}[-]\) denote images under the two sum constructors.  Then
\begin{align*}
  &\Delta(\operatorname{sumWord}(x_1,x_2),
          \operatorname{sumWord}(y_1,y_2)) \\
  &\quad=
    \mathsf{inl}[\Delta(x_1,y_1)]
    \cup
    \mathsf{inr}[\Delta(x_2,y_2)].
\end{align*}
The two image sets are disjoint because \(\mathsf{inl}\) and \(\mathsf{inr}\) are disjoint constructors.  Applying the inclusion-exclusion rule for disjoint finite sets reduces the cardinality of the union to the sum of the two cardinalities.  Since the left image is in bijection (via \(\mathsf{inl}\)) with the set of differing positions in \(x_1\) and \(y_1\), and similarly for the right image, the cardinalities equal \(\distH(x_1,y_1)\) and \(\distH(x_2,y_2)\) respectively, which gives the stated identity.

In the Lean proof, the filter decomposition is established by a finset extensionality argument using case analysis on \(\mathsf{inl}/\mathsf{inr}\).  Disjointness is proved by noting that an element cannot simultaneously be in the image of \(\mathsf{inl}\) and the image of \(\mathsf{inr}\).  The cardinality of each image under an injective map equals the cardinality of the source, which is recorded by
\[
  \texttt{Finset.card\_image\_of\_injective}.
\]
Together, these steps give the distance equality in a short \texttt{rw} chain.

\subsection{The product code}\label{subsec:product-code}

Given two finite codes \(C_1 \subseteq \mathtt{Word}\ \iota\ \beta_1\) and \(C_2 \subseteq \mathtt{Word}\ \kappa\ \beta_2\), their product code is the set of all concatenations:
\[
  C_1 \times C_2 := \{\,\operatorname{sumWord}(c_1,c_2) : c_1\in C_1,\ c_2\in C_2\,\}.
\]
In Lean, the product code is defined as the image of the Cartesian product under \(\operatorname{sumWord}\):
\begin{lstlisting}
def (*@\leanref{productCode}{CoveringCodes/Constructions/Product.lean}{74}@*) (C1 : Finset (Word iota beta1)) (C2 : Finset (Word kappa beta2)) :
    Finset (Word (Sum iota kappa) (Sum.elim beta1 beta2)) :=
  (C1 ×ˢ C2).image (fun p => sumWord p.1 p.2)
\end{lstlisting}
The expression \(\mathtt{C1} \times^{\mathsf{s}} \mathtt{C2}\) denotes the finset Cartesian product, whose elements are pairs \((c_1,c_2)\) with \(c_1\in C_1\) and \(c_2\in C_2\).  The \texttt{image} operation applies \(\operatorname{sumWord}\) to each pair, discarding duplicates.  The resulting finset has cardinality at most \(|C_1|\cdot|C_2|\), with equality when the mapping is injective on the Cartesian product.

In the \qary{} specialization, product codes are ordinary concatenations: a code
in length \(n_1\) and a code in length \(n_2\) embed into length \(n_1+n_2\)
via the standard bijection between the sum of coordinate types and
\(\Fin(n_1+n_2)\).  The formalization works with the sum-type coordinate index
directly, avoiding the need to state or apply any explicit bijection.

\subsection{Product covering theorem}\label{subsec:product-covering}

The main result of this section is the following.

\textbf{Product covering theorem.}\label{thm:product-covering}
Let \(C_1\) be a finite radius-\(r_1\) covering code over \(\iota\) and let \(C_2\) be a finite radius-\(r_2\) covering code over \(\kappa\).  Then the product code \(C_1\times C_2\) is a radius-\((r_1+r_2)\) covering code over \(\iota\oplus\kappa\).

The Lean statement is:
\begin{lstlisting}
theorem (*@\leanref{productCode_covers}{CoveringCodes/Constructions/Product.lean}{80}@*)
    (C1 : Finset (Word iota beta1)) (C2 : Finset (Word kappa beta2))
    (r1 r2 : Nat)
    (h1 : CoversFinset C1 r1) (h2 : CoversFinset C2 r2) :
    CoversFinset (productCode C1 C2) (r1 + r2)
\end{lstlisting}

\textbf{Proof.}  Let \(x : (s : \iota\oplus\kappa) \to (\beta_1\oplus\beta_2)(s)\) be an arbitrary word.  Define
\[
  x_1(i) := x(\mathsf{inl}\;i)
  \qquad\text{and}\qquad
  x_2(j) := x(\mathsf{inr}\;j).
\]
The reassembly identity gives \(x = \operatorname{sumWord}(x_1,x_2)\).  Since \(C_1\) covers at radius \(r_1\), there exists \(c_1\in C_1\) with \(\distH(x_1,c_1)\le r_1\).  Since \(C_2\) covers at radius \(r_2\), there exists \(c_2\in C_2\) with \(\distH(x_2,c_2)\le r_2\).

The product codeword \(\operatorname{sumWord}(c_1,c_2)\) belongs to \(C_1\times C_2\) because \((c_1,c_2)\in C_1\times^{\mathsf{s}} C_2\) and the product code is its image.  By the distance-additivity lemma,
\begin{align*}
  &\distH\!\bigl(x,\operatorname{sumWord}(c_1,c_2)\bigr) \\
  &\quad= \distH\!\bigl(\operatorname{sumWord}(x_1,x_2),\operatorname{sumWord}(c_1,c_2)\bigr) \\
  &\quad= \distH(x_1,c_1)+\distH(x_2,c_2) \le r_1+r_2.
\end{align*}
Since \(x\) was arbitrary, the product code covers at radius \(r_1+r_2\).

The Lean proof follows this argument step by step.  The decomposition of \(x\) is done by defining \(x_1\) and \(x_2\) as local let-bindings and proving the reassembly identity by \texttt{funext} with a case split on the sum type.  Membership in the product code is reduced to the standard image/product membership lemmas.  The distance bound uses the sum-word distance lemma and \texttt{Nat.add\_le\_add}.

\textbf{Remark on tightness.}  The radius \(r_1+r_2\) in Theorem~\ref{thm:product-covering} is in general not tight.  The product code has size at most \(|C_1|\cdot|C_2|\) and radius \(r_1+r_2\); it is not generally optimal for the combined space and radius.  The theorem is therefore most useful as a composition tool, not as a direct method for constructing optimal codes.

This product construction is the representative fully general construction in
the paper: define an operation on words, prove its distance behavior, turn it
into a finset transformation, and derive the corresponding covering theorem.
The next section uses the same certificate-oriented viewpoint for standard
relations between neighboring covering-code parameters.

\section{Neighbor Relations and Structural Code Transformations}\label{sec:neighbor-relations}

Many standard covering-code arguments are relations between neighboring
parameter triples.  A proof of \(K_q(n,r)\le U\) or \(K_q(n,r)\ge L\) can often
be transported to a proof about a different triple \((q',n',r')\).  For example,
puncturing a length-eight ternary radius-one cover gives a length-seven
radius-one cover of no larger size: extend a target word by one arbitrary
coordinate, use the original cover, and delete the added coordinate again.

The formalization packages such arguments as proof-carrying transformations of
the linked \leansource{upper-bound}{CoveringCodes/CoveringNumber.lean}{54} and
\leansource{lower-bound}{CoveringCodes/CoveringNumber.lean}{68} predicates.  The important
point is not the Lean syntax of each declaration, but the certified effect:
applying a relation theorem produces a new Lean proof that the transported
number is a valid upper or lower bound.  Table~\ref{tab:relation-rules}
summarizes the relation layer.  The source links point to representative
declarations; Appendix~\ref{app:artifact-map} gives a broader declaration map.

\begin{table*}[t]
\centering
\scriptsize
\setlength{\tabcolsep}{3pt}
\begin{tabular}{@{}p{0.19\textwidth}p{0.30\textwidth}p{0.34\textwidth}p{0.13\textwidth}@{}}
\toprule
Rule family & Certified effect & Proof idea & Source \\
\midrule
Radius monotonicity &
\(K_q(n,r)\le U \Rightarrow K_q(n,r+s)\le U\);
lower bounds move back from \(r+s\) to \(r\). &
Hamming balls grow with the radius. &
\leansource{radius rules}{CoveringCodes/Relations/RadiusMono.lean}{14} \\

Puncturing &
\(K_q(n,r)\le U \Rightarrow K_q(n-t,r)\le U\), with a lower-bound converse. &
Delete \(t\) coordinates; for the upper bound, fill deleted coordinates before
using the original cover. &
\leansource{length rules}{CoveringCodes/Relations/LengthTransforms.lean}{82} \\

Free lengthening &
\(K_q(n,r)\le U \Rightarrow K_q(n+t,r)\le q^tU\), with lower bound divided by
\(q^t\) using ceiling division. &
Append all possible suffixes to each codeword. &
\leansource{free lengthening}{CoveringCodes/Relations/LengthTransforms.lean}{123} \\

Dummy lengthening &
\(K_q(n,r)\le U \Rightarrow K_q(n+t,r+t)\le U\), with a lower-bound converse. &
Append fixed coordinates and spend at most one radius unit per new coordinate. &
\leansource{dummy lengthening}{CoveringCodes/Relations/LengthTransforms.lean}{52} \\

Alphabet projection and expansion &
Project from \(q\) symbols to \(a\le q\), or expand from \(q\) symbols to
\(Q\le sq\) with a factor \(s^n\). &
Use coordinatewise maps between finite alphabets and compare distances. &
\leansource{alphabet rules}{CoveringCodes/Relations/AlphabetTransforms.lean}{45} \\

Direct and repeated products &
Upper bounds multiply and radii add; repeated products give
\(K_q(mn,mr)\le U^m\). &
Specialize Section~\ref{subsec:product-covering} to \qary{} coordinate blocks:
concatenate independently covered coordinate blocks. &
\leansource{product rules}{CoveringCodes/Relations/DirectProduct.lean}{88} \\

Block grouping and ungrouping &
Group \(m\) \qary{} coordinates into one \(q^m\)-ary coordinate, or ungroup with
radius multiplied by \(m\). &
Compare block Hamming distance with coordinate Hamming distance. &
\leansource{block rules}{CoveringCodes/Relations/BlockTransforms.lean}{105} \\

Concatenation &
An outer \(Q\)-ary cover and an indexed \qary{} inner cover give a \qary{} cover of
length \(mN\) and radius \(N\rho+R(m-\rho)\). &
Approximate every target block by an inner symbol, then use the outer cover to
control the blocks whose chosen symbols differ. &
\leansource{concatenation}{CoveringCodes/Relations/Concatenation.lean}{147} \\

Structural shortening &
Explicit missing-symbol, fixed-coordinate, or pattern-avoidance information can
reduce both length and radius. &
A forced disagreement consumes radius before the shortened coordinates are
considered. &
\leansource{shortening}{CoveringCodes/Relations/StructuralShortening.lean}{21} \\

Hole filling &
A code covering at radius \(r\), plus a cover of its deep holes at radius
\(r-1\), gives a smaller-radius cover. &
Use the union of the original code and the auxiliary hole-covering code. &
\leansource{hole filling}{CoveringCodes/Relations/HoleFilling.lean}{41} \\

Colored product &
Matched color classes in two codes give a product code using only same-color
pairs. &
For every target pair, choose one color that supplies both approximations. &
\leansource{colored product}{CoveringCodes/Relations/ColoredProduct.lean}{72} \\
\bottomrule
\end{tabular}
\caption{Certified relation rules for transporting covering-code bounds.  The
table states the mathematical effect; each linked source declaration produces
the corresponding Lean proof in the certificate interface.}
\label{tab:relation-rules}
\end{table*}

The first seven rows are numerical relations: they transform only upper- or
lower-bound certificates.  These are the rules used by the database closure in
Section~\ref{sec:computation}.  The remaining rows are structural rules.  They
need an explicit code together with side information about that code, such as a
missing symbol, a forbidden pattern, a cover of deep holes, or a compatible
coloring.  Such information cannot be recovered from a bare numerical inequality
\(K_q(n,r)\le U\).

\subsection{Indexed concatenation}
\label{subsec:general-concatenation}

The concatenation rule is the most informative numerical relation because its
statement records more than a size inequality.  It combines an outer cover over
an alphabet of size \(Q\) with an explicit map from the \(Q\) outer symbols to
\qary{} inner blocks of length \(m\).  A hypothesis says that every \qary{} block is
within distance \(\rho\) of some indexed inner block.  If the outer code covers
length \(N\) at radius \(R\), then replacing each outer symbol by the
corresponding inner block gives a \qary{} code of length \(mN\) and radius
\[
  N\rho + R(m-\rho).
\]
The term \(N\rho\) pays for approximating each target block by an inner symbol.
In the at most \(R\) blocks where the selected outer codeword differs from that
symbol word, the block cost can rise from \(\rho\) to at most \(m\), adding
\(R(m-\rho)\).

This theorem is intentionally stated with an explicit inner map rather than only
a numerical certificate such as \(K_q(m,\rho)\le Q\).  The numerical certificate
proves that some inner code exists, but concatenation needs a fixed indexing of
outer alphabet symbols by inner blocks.  This is a typical formalization point:
the certificate must carry the data that the construction actually uses, not
only the numerical bound suggested by the construction.

\subsection{Structural explicit-code rules}
\label{subsec:structural-shortening}

Structural rules work from concrete codes.  In the missing-symbol shortening
rule, for example, no codeword uses symbol \(a\) at coordinate \(j\).  Any target
with symbol \(a\) at \(j\) therefore disagrees with every codeword at that
coordinate before the remaining coordinates are considered.  Deleting coordinate
\(j\) removes this forced disagreement and reduces the covering radius by one.
The pattern-avoidance version applies the same idea to several coordinates at
once: if every codeword differs from a prescribed partial pattern in at least
\(\delta\) coordinates, those \(\delta\) units of distance can be removed from
the radius after deleting the pattern coordinates.

Hole filling is another structural rule.  If \(C\) already covers the space at
radius \(r\), the only obstruction to a radius-\((r-1)\) cover is the finite set
of words not covered by \(C\) at radius \(r-1\).  A second code \(D\) that covers
exactly those deep holes at radius \(r-1\) yields a radius-\((r-1)\) cover of
size at most \(|C|+|D|\).  Colored products likewise require side information:
for every target pair, the left and right approximations must be available with
one shared color, so that their product codeword belongs to the same-color
product.

The relation layer changes the role of a certified database entry.  A primitive
bound can serve not only as a final reported result, but also as a seed from
which further certified bounds are obtained by applying relation rules.  The
next section demonstrates how published explicit code data become primitive
upper-bound certificates; Section~\ref{sec:computation} then explains how such
certificates are registered and propagated by trace replay.

\section{Case Study: Formalizing Published Covering Codes}\label{sec:casestudy}

This section demonstrates the end-to-end workflow for formalizing explicit
covering codes from the published literature.  The endpoint of the section is a
collection of ordinary \Lean{} theorems of type
\leanref{QaryKUpper}{CoveringCodes/CoveringNumber.lean}{54}; their database
registration is described separately in Section~\ref{sec:computation}.

\subsection{Source paper}

Van Laarhoven, Aarts, van Lint, and Wille~\cite{vanlaarhoven1989football} proved three upper bounds for the football pool problem by exhibiting explicit covering codes:
\[
  K_3(6,1)\le 73, \qquad K_3(7,1)\le 186, \qquad K_3(8,1)\le 486.
\]
All three bounds were obtained using simulated annealing.  For $n=8$, the search was carried out in a Blokhuis--Lam style setup involving a small set \(S\) and a matrix \(A\); the authors then describe a systematic construction producing many distinct 486-element covering codes from that result.  The codes were published as printed grid figures in the paper.

\subsection{Data extraction}

The code data were transcribed from the published figures and the
Blokhuis--Lam style \(n=8\) construction in van Laarhoven et al. into packed
arrays in the \leansource{van Laarhoven source module}{CoveringCodes/Database/Sources/VanLaarhoven1989.lean}{230}.  This
transcription was an untrusted data-production step: it was AI-assisted and
manually corrected, and is not treated as a reproducible part of the artifact.
The checked artifact starts with the committed \Lean{} definitions of the
finite sets.  \Lean{} then verifies their cardinality and radius-one covering
property from finite certificates.  Thus the formal theorem depends on the
machine-checked covering certificate, not on trusting the transcription process.

\subsection{Lean encoding}

A naive encoding of 486 codewords as a finset literal hits Lean's elaboration recursion depth limit.  The solution is to represent each word as a single natural number via the ternary packing
\[
  \operatorname{enc}(w) = \sum_{i=0}^{n-1} w_i \cdot 3^i
\]
and store the codes as an \texttt{Array~\(\N\)}.  The covering finset is then
reconstructed by \leanref{codeFromPacked}{CoveringCodes/Database/Sources/VanLaarhoven1989.lean}{142}
from the packed arrays; the largest one begins at
\leanref{vanLaarhoven8Packed}{CoveringCodes/Database/Sources/VanLaarhoven1989.lean}{402}.
This
representation scales better for large explicit code data because it avoids
large finset literals and elaboration recursion limits.

The covering proof does not ask Lean to search for a nearby codeword from
scratch.  For each ambient word, the source stores a compact move certificate:
either keep the word itself, or change one coordinate to one ternary value.
Lean checks that the suggested move produces a packed word contained in the
code.  A \leansource{checked lemma}{CoveringCodes/Database/Sources/VanLaarhoven1989.lean}{218}
then turns the Boolean certificate pass into a
\leanref{CoversFinset}{CoveringCodes/Covers.lean}{26} proof.  The cardinality proof is separate and uses the
array length.  The public declarations listed in
Appendix~\ref{app:artifact-map} package the resulting finite sets, cardinality
bounds, and covering proofs.

\subsection{Switchable finite proof mode and resource use}
\label{subsec:proof-mode-resources}

The linked \leansource{wrapper tactic}{CoveringCodes/Database/ProofMode.lean}{16}
closes the finite Boolean checks.
By default it calls ordinary \texttt{decide}, so the resulting proof is replayed
by Lean's kernel reduction path.  Setting the proof-mode option to native mode
switches the same proof leaves to \texttt{native\_decide}.  This is useful for
development and measurement, but it uses Lean's native compiled evaluator and
therefore the trust boundary of Section~\ref{subsec:db-trust-boundary}.

Table~\ref{tab:proof-mode-resources} reports file-level compile measurements for
three expensive source files.  Each file was checked by invoking Lean through
Lake with one worker, a 400 GiB memory cap, profiling enabled, and the proof-mode
option set either to kernel mode or to native mode.  The runs were performed on
a server with two Intel Xeon Gold 6438M sockets, 64 physical cores / 128
hardware threads, and 512 GiB of RAM.  All six runs completed successfully with
exit code 0 and no timeout.  The numbers are wall-clock and maximum resident set
size for whole-file checks, not per-theorem attributions.

\begin{table*}[t]
\centering
\footnotesize
\begin{tabular}{@{}lrrrr@{}}
\toprule
Source file & \multicolumn{2}{c}{Kernel mode} & \multicolumn{2}{c}{Native mode} \\
\cmidrule(lr){2-3}\cmidrule(l){4-5}
 & Wall time & Max RSS & Wall time & Max RSS \\
\midrule
\texttt{SmallExplicitUpper.lean} & 99.5 min & 58.7 GiB & 24.0 s & 3.76 GiB \\
\texttt{SmallLowerBounds.lean} & 103.1 s & 7.46 GiB & 57.1 s & 3.63 GiB \\
\texttt{VanLaarhoven1989.lean} & 77.1 min & 357.0 GiB & 37.7 s & 3.75 GiB \\
\bottomrule
\end{tabular}
\caption{File-level resource measurements for the switchable finite proof mode.
Kernel mode uses ordinary \texttt{decide}; native mode uses
\texttt{native\_decide}.  The van Laarhoven row aggregates the
\(K_3(6,1)\), \(K_3(7,1)\), and \(K_3(8,1)\) certificates in one source file.}
\label{tab:proof-mode-resources}
\end{table*}

These measurements explain why both modes are useful.  Kernel mode gives the
stronger proof-replay story, but it can require hours and hundreds of GiB for
the largest file-level checks in the present artifact.  Native mode keeps the
same finite certificate structure and is practical for routine development, but
uses the native-evaluation trust boundary.
The wall-clock times in Table~\ref{tab:proof-mode-resources} should be read as
representative measurements rather than deterministic values; reruns on the same
server can vary by several minutes for the longest kernel checks and by a few
seconds for native checks.

\subsection{Certificate packaging}\label{subsec:casestudy-certificate}

Each explicit-code record bundles a finset code with its cardinality bound and
covering proof.  The record projection
\leanref{ExplicitQaryUpper.toUpper}{CoveringCodes/Database/ExplicitCode.lean}{14}
extracts the corresponding
\leanref{QaryKUpper}{CoveringCodes/CoveringNumber.lean}{54} proof term.  This completes the code-specific
part of the formalization: the data extracted from the published figures has
been reconstructed as a finite set of \qary{} words and turned into
machine-checked upper-bound certificates.  The formal theorem asserts
\emph{at most} 486 codewords for the largest case; it does not certify
row-for-row identity with the printed grid.

\subsection{What this demonstrates}

Three properties of the formalization workflow are illustrated by this case study.

\textbf{Independent verification of published results.}  The extraction step detected errors in the initial transcription of the printed figures.  Machine verification provides a check on published results that are rarely read in full detail by later readers.

\textbf{Scalable Lean encoding.}  Packing codewords as natural numbers and
reconstructing the finset by unpacking the stored array entries and applying
\texttt{List.toFinset} avoids large finset literals and keeps the source file
manageable.  The same encoding can be reused
for larger explicit \qary{} codes, but it only addresses source-size and
elaboration bottlenecks; the covering check itself still grows with the ambient
space and the number of codewords.

\textbf{Finite proof modes.}  Computational upper-bound certificates reduce to a
covering check over a finite domain, which is decidable.  The artifact source
can replay the expensive finite proof leaves either through Lean's ordinary
\texttt{decide} path or, when the proof-mode option is enabled, through
\texttt{native\_decide}.  The latter is much faster for routine development, but
it adds the trust assumption of Lean's native compiled evaluation path.  The
general pattern is to keep the external search or transcription untrusted while
Lean checks the resulting explicit certificate: extract the codewords, pack
them, verify the finite covering predicate, and wrap the result as an
\leanref{ExplicitQaryUpper}{CoveringCodes/Database/ExplicitCode.lean}{7}.  This should not be read as an automatic method for
arbitrarily large covering-code instances.  A lower bound from an exhaustive
non-existence argument would require its own certificate format or formal
argument and is future work rather than a consequence of this upper-bound
wrapper.

\section{Certified Bounds and Proof-Carrying Database}\label{sec:computation}

The preceding case study produced reusable Lean certificates from published
covering codes.  This section explains how such certificates are combined with
formal relation rules to obtain further certified bounds.  The point of the
database is not merely to store numerical intervals.  Each stored upper or lower
bound carries a derivation trace whose replay constructs a Lean proof of the
corresponding \leanref{QaryKUpper}{CoveringCodes/CoveringNumber.lean}{54} or
\leanref{QaryKLower}{CoveringCodes/CoveringNumber.lean}{68} statement.

\subsection{Certificate structure}\label{subsec:db-architecture}

The database uses the same upper/lower certificate interface as the rest of the
paper.  Primitive leaves are ordinary Lean theorems: explicit codes give
\leanref{QaryKUpper}{CoveringCodes/CoveringNumber.lean}{54} proofs, while counting or non-existence arguments give
\leanref{QaryKLower}{CoveringCodes/CoveringNumber.lean}{68} proofs.  Relation rules from
Section~\ref{sec:neighbor-relations} transport such proofs between neighboring
parameter triples.  The trace layer records the sequence of primitive leaves
and relation applications; replaying the trace gives the proof object used by
the public query result.

Table~\ref{tab:db-certificate-map} gives the main declaration-level audit map
for this certificate layer.  A broader artifact map is given in
Appendix~\ref{app:artifact-map}.

\begin{table*}[t]
\centering
\scriptsize
\begin{tabular}{@{}p{0.25\textwidth}p{0.27\textwidth}p{0.29\textwidth}p{0.13\textwidth}@{}}
\toprule
Claim & Lean declarations & File & Trust boundary \\
\midrule
Upper, lower, and exact covering-number certificates &
\leanref{KUpper}{CoveringCodes/CoveringNumber.lean}{50},
\leanref{KLower}{CoveringCodes/CoveringNumber.lean}{64},
\leanref{KSpec}{CoveringCodes/CoveringNumber.lean}{33},
\leanref{KSpec.ofUpperLower}{CoveringCodes/CoveringNumber.lean}{89} &
\leanfileat{CoveringCodes/CoveringNumber.lean}{33} &
Handwritten Lean; kernel checked. \\

Trace replay constructs bound proofs &
\leanref{UpperTrace.valid}{CoveringCodes/Database/Trace.lean}{108},
\leanref{LowerTrace.valid}{CoveringCodes/Database/Trace.lean}{123} &
\leanfileat{CoveringCodes/Database/Trace.lean}{10} &
Handwritten Lean; kernel checked. \\

Generated query results carry proof fields and are internally consistent &
\leanref{BestBounds}{CoveringCodes/Database/Defs.lean}{8},
\leanref{BestBounds.consistent}{CoveringCodes/Database/Defs.lean}{26},
\leanref{bestBounds}{CoveringCodes/Database/GeneratedAPI.lean}{15},
\leanref{bestBounds_consistent}{CoveringCodes/Database/GeneratedAPI.lean}{18} &
\leanfileat{CoveringCodes/Database/Defs.lean}{8},
\leanfileat{CoveringCodes/Database/GeneratedAPI.lean}{15} &
Generated traces plus checked replay. \\

Explicit finite codes become upper-bound certificates &
\leanref{ExplicitQaryUpper}{CoveringCodes/Database/ExplicitCode.lean}{7},
\leanref{ExplicitQaryUpper.toUpper}{CoveringCodes/Database/ExplicitCode.lean}{14} &
\leanfileat{CoveringCodes/Database/ExplicitCode.lean}{7} &
Finite checks use the proof mode stated in Section~\ref{subsec:db-trust-boundary}. \\

van Laarhoven bounds enter as checked upper-bound theorems &
\leanref{vanLaarhoven6Upper_valid}{CoveringCodes/Database/Sources/VanLaarhoven1989.lean}{288},
\leanref{vanLaarhoven7Upper_valid}{CoveringCodes/Database/Sources/VanLaarhoven1989.lean}{393},
\leanref{vanLaarhoven8Upper_valid}{CoveringCodes/Database/Sources/VanLaarhoven1989.lean}{635} &
\leanfileat{CoveringCodes/Database/Sources/VanLaarhoven1989.lean}{282} &
Committed data checked by Lean; transcription is untrusted. \\
\bottomrule
\end{tabular}
\caption{Audit map for the proof-carrying database interface.}
\label{tab:db-certificate-map}
\end{table*}

\leansource{The precomputed table}{reference-data/lean/non_mixed_covering_codes.csv}{1}
is an implementation of this certificate structure over the parameter box
\(0\le q\le21\), \(0\le n\le48\), and \(0\le r\le48\).  It contains 52,822
entries serialized as Lean source in 212 generated chunks.  The
generator that writes these chunks is not verified.  Its output is accepted as
part of the artifact only because Lean type-checks the generated trace values
and the public query interface
\leanref{bestBounds}{CoveringCodes/Database/GeneratedAPI.lean}{15} obtains proof
fields by trace replay.

\subsection{Trust boundary}\label{subsec:db-trust-boundary}

The formalization separates checked proof objects from the computations and
scripts that propose them.

\begin{itemize}
  \item Handwritten mathematical theorems, including the ball-volume formula,
        sphere-covering lower bound, product construction, small lower-bound
        arguments, and relation theorems, are ordinary Lean declarations checked
        by the kernel.
  \item Generated database entries are Lean source code containing explicit
        \leanref{LowerTrace}{CoveringCodes/Database/Trace.lean}{55} and
        \leanref{UpperTrace}{CoveringCodes/Database/Trace.lean}{10} values.  The generator is
        not itself verified; the generated entries are checked by Lean through
        trace typing and replay.
  \item External transcription and generation steps, including the transcription
        for the van Laarhoven case study and the table generator, are untrusted
        producers of candidate data or Lean source.  Their output contributes to
        a theorem only after the resulting Lean declarations are accepted.
  \item Goals closed with \texttt{native\_decide} use Lean's native compiled
        evaluator rather than kernel normalization for the finite decision
        procedure.  The source supports both ordinary \texttt{decide} replay
        and this faster native mode for expensive finite leaves; only native
        proof-mode builds use this additional trust assumption.
\end{itemize}

Thus the database is proof-carrying with respect to Lean trace replay, while
the computations that found, generated, or accelerated some leaves remain part
of the stated trusted infrastructure.

\subsection{Imported explicit codes}
\label{subsec:db-vlaarhoven-integration}

The van Laarhoven certificates from
Section~\ref{subsec:casestudy-certificate} enter the database as primitive
upper-bound leaves.  The checked declarations
\leansource{\(n=6\)}{CoveringCodes/Database/Sources/VanLaarhoven1989.lean}{288},
\leansource{\(n=7\)}{CoveringCodes/Database/Sources/VanLaarhoven1989.lean}{393},
and
\leansource{\(n=8\)}{CoveringCodes/Database/Sources/VanLaarhoven1989.lean}{635}
prove \(K_3(6,1)\le73\), \(K_3(7,1)\le186\), and \(K_3(8,1)\le486\)
in the certificate interface.
Once registered as primitive sources, these theorems are propagated only through
the generic relation rules; there is no van-Laarhoven-specific recursion in the
closure layer.

\subsection{A two-block cyclic upper bound for \texorpdfstring{\(K_q(3,1)\)}{Kq(3,1)}}
\label{subsec:db-two-block-cyclic}

One primitive upper source deserves separate mention because it is a uniform
construction in \(q\), not a list of isolated finite codes.  Split the alphabet
\(\{0,\ldots,q-1\}\) into two consecutive blocks of sizes
\[
  a=\lfloor q/2\rfloor,\qquad b=q-a=\lceil q/2\rceil .
\]
For a nonempty block of size \(m\), index the block by
\(\mathbb{Z}/m\mathbb{Z}\) and use the \(m^2\) length-three codewords
\[
  (i,\; i+j,\; j) \qquad (i,j\in \mathbb{Z}/m\mathbb{Z}),
\]
with addition modulo \(m\).  Taking the union of this cyclic code in the two
blocks gives at most \(a^2+b^2\) codewords.

The covering argument is a small pigeonhole argument.  Any target word
\((x_0,x_1,x_2)\) has at least two coordinates in the same block.  If these are
positions \(0\) and \(1\), choose \(i=x_0\) and \(j=x_1-x_0\) inside that
block; if they are positions \(0\) and \(2\), choose \(i=x_0\) and \(j=x_2\);
and if they are positions \(1\) and \(2\), choose \(j=x_2\) and
\(i=x_1-x_2\).  In each case the corresponding cyclic codeword agrees with
the target in two coordinates and is therefore within Hamming distance one.

The Lean theorem
\leanref{twoBlockCyclic331Upper_valid}{CoveringCodes/Database/Sources/TwoBlockCyclic.lean}{249}
proves, for every \(q\), the certificate
\[
  K_q(3,1)\le (q/2)^2+(q-q/2)^2 .
\]
Here \(q/2\) is Lean's natural-number floor division, so the two block sizes are
\(\lfloor q/2\rfloor\) and \(\lceil q/2\rceil\).  Thus the right-hand side is
\(\lceil q^2/2\rceil\).  The artifact proves this as an upper bound for every
\(q\).  It does not prove the corresponding lower bound, so it does not certify
the exact value of \(K_q(3,1)\) for all \(q\).  The comparison table agrees with
this value for \(2\le q\le21\), but that table is used only as untrusted
comparison data, not as a proof input.

\subsection{Selected database certificates}
\label{subsec:db-selected-exact}

The database contains selected exact covering-number certificates beyond the
structural families of radius zero, large radius, and near diameter, and it also
reports non-exact certified intervals.  These entries were useful development
targets because they exercise both sides of the certificate interface: an
explicit covering code or construction for the upper bound, and an independent
lower-bound proof.  Table~\ref{tab:selected-db-certificates} lists a compact
sample of representative query results.  Exact rows are written with \(=\), while
non-exact interval rows are written with \(\in [L,U]\).  The examples are not
an exhaustive inventory of the database; they were chosen to show exactness
certificates, sparse-slice lower bounds, explicit upper-bound seeds, and
automatic closure propagation.

\begin{table*}[t]
\centering
\scriptsize
\begin{tabular}{@{}>{\raggedleft\arraybackslash}p{0.28\textwidth}@{\hspace{1.0em}}c@{\hspace{1.0em}}>{\raggedright\arraybackslash}p{0.50\textwidth}@{}}
\toprule
Lower certificate & Certified statement & Upper certificate \\
\midrule
sphere-covering bound & \(K_2(3,1)=2\) & constant-symbol code, \(K_q(n,r) \le q\) when \(q\cdot(n-r-1) < n\) \\
binary parity-hole proof & \(K_2(5,1)=7\) & explicit binary radius-one code \\
binary normalized-hole proof & \(K_2(6,2)=4\) & explicit binary radius-two code \\
sphere-covering bound & \(K_2(7,1)=16\) & binary Hamming code \\
ternary overlap-counting proof & \(K_3(3,1)=5\) & explicit ternary radius-one code \\
sphere-covering bound & \(K_3(4,1)=9\) & ternary Hamming code \\
sparse-slice proof for \(K_q(3,1)\) & \(K_4(3,1)=8\) & explicit quaternary radius-one code \\
sphere-covering bound & \(K_4(5,1)=64\) & quaternary Hamming radius-one code \\
sphere-covering bound & \(K_5(3,1)\in[10,13]\) & explicit quinary radius-one code \\
sphere-covering bound & \(K_6(4,1)\in[62,108]\) & explicit senary radius-one seed propagated by free lengthening \(\times q^1\) \\
sparse-slice proof for \(K_q(3,1)\) & \(K_9(3,1)\in[36,41]\) & explicit nonary radius-one code \\
\bottomrule
\end{tabular}
\caption{Selected exact and interval certificates reported by the generated
database.  External tables were used only as untrusted target-selection aids;
the statements shown here are certified by Lean proofs and database trace
replay, but the intervals are not claimed to be best known values.}
\label{tab:selected-db-certificates}
\end{table*}

The concrete finite-code upper-bound seeds shown in
Table~\ref{tab:selected-db-certificates} are checked by ordinary \texttt{decide};
they do not rely on \texttt{native\_decide}.  Sphere-covering exact rows pair
those upper bounds with the formal ball-volume lower bound.
The \leansource{parity-hole proof}{CoveringCodes/Database/Sources/SmallLowerBounds.lean}{583}
for \(K_2(5,1)\) splits a hypothetical
six-word radius-one cover into even and odd centers and derives an uncovered
parity class after a small finite check.  The
\leansource{normalized-hole proof}{CoveringCodes/Database/Sources/SmallLowerBounds.lean}{664}
for
\(K_2(6,2)\) translates any three centers so that one is zero, then checks that
two remaining centers cannot cover all length-six binary words at radius two.
For \(K_3(3,1)\), the \leansource{overlap-counting lower-bound theorem}{CoveringCodes/Database/Sources/SmallLowerBounds.lean}{171}
formalizes the overlap count showing that four ternary radius-one balls in
length three cover at most 26 of the 27 words.  For \(K_4(3,1)\), the
\leansource{sparse-slice lower bound}{CoveringCodes/Database/Sources/SparseSlicer.lean}{201}
specializes a general length-three radius-one
argument: a sparse first-coordinate fiber forces many distinct last-two
coordinate projections among the codewords.

The provenance labels in the database name these proof mechanisms rather than
external table references.  This is intentional.  External tables and local
searches are useful for finding candidate triples and candidate codes, but the
trusted artifact begins with the Lean definitions, the checked finite covering
predicates, and the lower-bound theorems registered as primitive sources.

\subsection{Closure and query consistency}
\label{subsec:db-closure}
\label{subsec:db-query}

The closure layer applies the certified relation rules to primitive sources
until no bound in the bounded parameter box improves.  The current generated
table records 52,822 entries.  A generated entry stores lower and upper traces,
not just two natural numbers.  When the public query
\leanref{bestBounds}{CoveringCodes/Database/GeneratedAPI.lean}{15} returns a
\leanref{BestBounds}{CoveringCodes/Database/Defs.lean}{8} record, its
\texttt{lower\_proof} and \texttt{upper\_proof} fields are obtained by replaying
those traces.

The proof of consistency is mathematical rather than an invariant of the table
generator.  The upper proof supplies a covering code \(C\) with
\(|C|\le\texttt{upper}\).  The lower proof applies to this same covering code
and gives \(\texttt{lower}\le |C|\).  Transitivity gives
\(\texttt{lower}\le\texttt{upper}\).  Appendix~\ref{app:closure-diagnostics}
records one generated-table run and the resulting trace-constructor counts.

\subsection{Path toward solver certificates}\label{subsec:db-solver}

The same certificate interface can host computational certificates.  An upper
bound from a SAT or integer-programming search can enter as an explicit code
once Lean checks that the code covers the ambient space.  A lower bound from an
exhaustive non-existence search would enter as a named \texttt{QaryKLower}
theorem, possibly through a dedicated certificate wrapper.  In both cases the
database stores the resulting theorem as a primitive leaf and lets the certified
relation rules propagate any derived bounds.

\section{Prior Art}\label{sec:prior-art}

The mathematical theory underlying this work is classical.  The monograph of
Cohen, Honkala, Litsyn, and Lobstein remains the standard reference for covering
codes, including basic bounds, constructions, puncturing arguments, products,
and tables of known values~\cite{cohen1997covering}.  Further work on \qary{}
upper bounds, the football-pool problem, and ternary covering codes provides
many of the concrete constructions and numerical targets that motivate this
formalization~\cite{ostergard1991upper,hamalainen1995football,habsieger1996ternary,vanlaarhoven1989football}.
Recent work of Gijswijt and Polak strengthens semidefinite lower-bound methods
for covering codes~\cite{gijswijt2025sdp}.

The computational prior art is also substantial.  Tables such as Keri's database
record best known bounds for many triples $(q,n,r)$ and are indispensable source
material for new formal certificates~\cite{keriCoveringCodes}.  These tables and
the papers behind them usually provide numerical bounds, explicit code data,
program output, or mathematical proofs.  They do not, however, provide proof
objects that can be checked by a theorem prover and composed with other certified
bounds.

\subsection{Related formalization work}
\label{subsec:related-formalization}

Coding theory has been formalized before, most substantially in the
\textsf{InfoTheo} library for the Rocq/Coq proof-assistant
ecosystem~\cite{affeldt2020linear}.  That work develops a broad
information-theoretic and coding-theoretic foundation, including linear
error-correcting codes, the Hamming metric, minimum distance, and families such
as Hamming, Reed--Solomon, and BCH codes, together with LDPC decoder material.
The emphasis of that library is the error-correcting setting: encoding,
decoding, minimum distance, and associated bounds.

Existing proof-assistant formalizations of coding theory mainly develop the
error-correcting, or packing, side of the subject: linear codes, minimum
distance, decoding, and classical code families.  The present development
formalizes the covering side, where the central object is \(K_q(n,r)\) and the
main proof objects are upper-, lower-, and exactness certificates for finite
covers.  We are not aware of previous formalizations of \qary{} covering
numbers, the sphere-covering lower bound, or proof-carrying covering-code
databases.  The library reuses finite-type and Hamming-distance infrastructure
where possible, but the covering predicates, certificate interface, and trace
replay layer are specific to this development.  The \qary{} specialization uses
\(\Fin n\to\Fin q\) for arbitrary \(q\), not only finite-field alphabets.

The contribution is also partly architectural.  The formalization separates
upper-bound, lower-bound, and exactness certificates into independent
predicates (\leanref{KUpper}{CoveringCodes/CoveringNumber.lean}{50},
\leanref{KLower}{CoveringCodes/CoveringNumber.lean}{64},
\leanref{KSpec}{CoveringCodes/CoveringNumber.lean}{33}) and connects
concrete bounds through a trace-carrying database whose entries replay to Lean
proofs.  This design is intended to make later computational certificates,
including exhaustive-search, integer-programming, or semidefinite-programming
certificates, attachable to the same interface without changing the trusted
core.  The present development is written in the style of \mathlib{} and reuses
its Hamming-distance and finite-type infrastructure~\cite{mathlib2020}.  It is
not integrated into \mathlib{}, but the definitions are designed to be
compatible with eventual upstreaming.

The contribution of this paper is therefore not a new theoretical bound.  It is
a formal interface that turns standard covering-code ingredients into reusable
Lean certificates: explicit upper-bound witnesses, lower-bound predicates,
exactness statements, certified relation rules, and trace-carrying database
entries.  In this sense the work sits between the classical covering-code
literature and future computational certificates: it formalizes enough of the
theory to make published or solver-generated covering-code evidence checkable
inside Lean.

\section{Discussion and Future Work}\label{sec:future-work}

This paper formalizes a foundation rather than a large table of new
covering-code bounds.  The main design choice is to separate mathematical claims
about \(K_q(n,r)\) into upper-bound certificates, lower-bound certificates, and
exactness certificates.  This separation proved useful throughout the
development: explicit codes, counting arguments, elementary exact values, and
database entries can be checked independently and then recombined when their
numerical bounds meet.

Several formalization lessons emerged from this choice.  First, treating
\(K_q(n,r)\) through certificate predicates rather than an immediate minimum
function kept upper-bound witnesses and lower-bound arguments modular.  This
matters in the selected exact certificates, where explicit-code upper bounds and
independent lower-bound proofs meet only at \texttt{KSpec}.
Second, the representation \(\Fin n \to \Fin q\) made edge cases such as \(q=0\),
\(n=0\), and \(r\ge n\) part of the ordinary theorem statements rather than
external conventions.  For example, the formal statements distinguish the empty
positive-length word space over \(\Fin 0\) from the unique length-zero word, so
the boundary cases in Table~\ref{tab:edge-cases} are theorems rather than side
conventions.  Third, the trace layer showed a useful separation between
unverified producers of candidate data and checked Lean proof replay: new
sources can be added as local theorems while the database infrastructure
propagates them through certified relation rules.  In particular, the table
generator and the van Laarhoven transcription propose data, but a database entry
contributes to a paper claim only after its trace or explicit-code wrapper
produces a Lean proof.

The current formalization is intentionally conservative.  It covers finite
Hamming spaces, closed balls, covering predicates, elementary exact cases, the
\qary{} ball-volume formula, the sphere-covering bound, a product construction,
and a substantial layer of certified neighbor relations.  These results are
standard, but their formal versions make all hypotheses and edge cases explicit.
In particular, cases such as \(q=0\), \(n=0\), and \(r\ge n\) are not handled by
informal convention; they are consequences of the definitions used in Lean.

The proof-carrying database is the other main outcome.  Its purpose is not
merely to store numbers, but to store bounds together with derivation traces that
replay to Lean proofs.  This gives a reproducibility property that ordinary
numerical tables usually do not provide: each reported database bound is
returned with a trace that Lean can replay to a proof of the corresponding
upper- or lower-bound predicate.  This should be understood in the trust model
of Section~\ref{subsec:db-trust-boundary}: the table generator is not verified,
and native-mode explicit-code leaves rely on Lean's
\leansource{native evaluation path}{CoveringCodes/Database/ProofMode.lean}{16}.
The current table is still small as mathematical data,
since it contains elementary sources, selected exact covering-number
certificates, table-guided upper-bound seeds, and the three van Laarhoven
upper-bound certificates, rather than a reproduction of the known covering-code
tables.  Its value is architectural: adding a new source theorem exposes it to
the existing closure machinery for the bounded parameter range currently
generated.

There are several natural directions for extending the formalization.  The first
is to add more of the classical covering-code lower-bound theory, beyond the
sphere-covering bound.  This includes stronger combinatorial bounds and, in the
longer term, LP or SDP bounds.  Such results
would require additional formal infrastructure, but they fit the existing
\leanref{QaryKLower}{CoveringCodes/CoveringNumber.lean}{68} interface.

A second direction is to import more explicit covering codes from the
literature.  The van Laarhoven case study shows the basic workflow for moderate
finite instances: extract the code data, encode it compactly, verify the finite
covering condition through the switchable proof mode, and register the resulting
theorem as a primitive source.  Scaling this process to substantially larger
codes would require attention to runtime, memory, and possibly more structured
certificate formats; it is not automatic from the present encoding alone.

Finally, the database closure itself can be strengthened.  The direct-product
rule is already formalized, but it is not used in the automatic fixpoint closure
because of its cost.  More selective search strategies, cached product
applications, or user-guided trace construction may make product-derived bounds
practical without making table generation too expensive.  Structural rules such
as shortening, hole filling, concatenation, and colored products also suggest a
future database layer that stores not only numerical bounds but explicit code
structure.

Overall, the formalization establishes a machine-checked base for \qary{} covering
codes in Lean.  The next step is not to replace the existing covering-code
literature, but to connect it to this base: published constructions,
computational searches, and stronger lower-bound arguments can be connected to
the same certificate interfaces once suitable Lean statements, certificate
formats, and trust boundaries are supplied.

\appendix

\section{Artifact and Declaration Map}\label{app:artifact-map}

Table~\ref{tab:artifact-map} maps the main paper-level claims to Lean
declarations and artifact files.  The links point to the fixed artifact commit
named in the artifact availability paragraph.  Build and regeneration
instructions are in \leanfileat{README.md}{13}; all paths are relative to the
repository root.

\begin{table*}[t]
\centering
\scriptsize
\setlength{\tabcolsep}{2pt}
\begin{tabular}{@{}p{0.24\textwidth}p{0.32\textwidth}p{0.27\textwidth}p{0.13\textwidth}@{}}
\toprule
Paper claim & Lean declarations & Artifact file & Trust boundary \\
\midrule
Finite Hamming spaces and \qary{} words &
\leanref{Word}{CoveringCodes/Basic.lean}{20},
\leanref{QaryWord}{CoveringCodes/Basic.lean}{24},
\leanref{dist}{CoveringCodes/Basic.lean}{44},
\leanref{qaryWord_card}{CoveringCodes/Basic.lean}{81} &
\leanfileat{CoveringCodes/Basic.lean}{20} &
Handwritten Lean; kernel checked. \\

Covering predicates and exactness certificates &
\leanref{Covers}{CoveringCodes/Covers.lean}{22},
\leanref{CoversFinset}{CoveringCodes/Covers.lean}{26},
\leanref{KUpper}{CoveringCodes/CoveringNumber.lean}{50},
\leanref{KLower}{CoveringCodes/CoveringNumber.lean}{64},
\leanref{KSpec}{CoveringCodes/CoveringNumber.lean}{33},
\leanref{KSpec.ofUpperLower}{CoveringCodes/CoveringNumber.lean}{89} &
\leanfileat{CoveringCodes/Covers.lean}{22},
\leanfileat{CoveringCodes/CoveringNumber.lean}{33} &
Handwritten Lean; kernel checked. \\

Structural exact cases &
\leanref{qaryKSpec_zero_radius}{CoveringCodes/SmallCases/ZeroRadius.lean}{53},
\leanref{singleton_covers_of_card_le}{CoveringCodes/SmallCases/LargeRadius.lean}{57},
\leanref{qaryKSpec_radius_nMinusOne}{CoveringCodes/SmallCases/RadiusNMinusOne.lean}{202} &
\leanfileat{CoveringCodes/SmallCases/ZeroRadius.lean}{53},
\leanfileat{CoveringCodes/SmallCases/LargeRadius.lean}{57},
\leanfileat{CoveringCodes/SmallCases/RadiusNMinusOne.lean}{202} &
Handwritten Lean; kernel checked. \\

Ball volume and sphere-covering lower bound &
\leanref{qaryBallVolume}{CoveringCodes/Bounds/Balls.lean}{42},
\leanref{qaryBallVolume_eq_card}{CoveringCodes/Bounds/Balls.lean}{269},
\leanref{sphere_covering_bound}{CoveringCodes/Bounds/SphereCoveringBound.lean}{40} &
\leanfileat{CoveringCodes/Bounds/Balls.lean}{42},
\leanfileat{CoveringCodes/Bounds/SphereCoveringBound.lean}{40} &
Handwritten Lean; kernel checked. \\

Product construction and relation rules &
\leanref{productCode_covers}{CoveringCodes/Constructions/Product.lean}{80},
\leanref{upper_radius_mono}{CoveringCodes/Relations/RadiusMono.lean}{14},
\leanref{upper_puncture}{CoveringCodes/Relations/LengthTransforms.lean}{82},
\leanref{upper_direct_product}{CoveringCodes/Relations/DirectProduct.lean}{88},
\leanref{upper_block_group}{CoveringCodes/Relations/BlockTransforms.lean}{105} &
\leanfileat{CoveringCodes/Constructions/Product.lean}{80},
\leandir{CoveringCodes/Relations/} &
Handwritten Lean; kernel checked. \\

Selected small exact certificates &
\leanref{smallLowerBinary251Lower_valid}{CoveringCodes/Database/Sources/SmallLowerBounds.lean}{583},
\leanref{smallLowerBinary262Lower_valid}{CoveringCodes/Database/Sources/SmallLowerBounds.lean}{664},
\leanref{smallLowerTernary331Lower_valid}{CoveringCodes/Database/Sources/SmallLowerBounds.lean}{171},
\leanref{sparseBound_valid}{CoveringCodes/Database/Sources/SparseSlicer.lean}{201} &
\leanfileat{CoveringCodes/Database/Sources/SmallLowerBounds.lean}{171},
\leanfileat{CoveringCodes/Database/Sources/SparseSlicer.lean}{201},
\leanfileat{CoveringCodes/Database/Sources/SmallExplicitUpper.lean}{259} &
Handwritten Lean plus finite checks; kernel checked unless native proof mode is selected. \\

Uniform two-block cyclic construction &
\leanref{twoBlockCyclic331Upper_valid}{CoveringCodes/Database/Sources/TwoBlockCyclic.lean}{249} &
\leanfileat{CoveringCodes/Database/Sources/TwoBlockCyclic.lean}{249} &
Handwritten Lean; kernel checked. \\

van Laarhoven upper-bound certificates &
\leanref{vanLaarhoven6Explicit}{CoveringCodes/Database/Sources/VanLaarhoven1989.lean}{282},
\leanref{vanLaarhoven7Explicit}{CoveringCodes/Database/Sources/VanLaarhoven1989.lean}{387},
\leanref{vanLaarhoven8Explicit}{CoveringCodes/Database/Sources/VanLaarhoven1989.lean}{629},
\leanref{vanLaarhoven6Upper_valid}{CoveringCodes/Database/Sources/VanLaarhoven1989.lean}{288},
\leanref{vanLaarhoven7Upper_valid}{CoveringCodes/Database/Sources/VanLaarhoven1989.lean}{393},
\leanref{vanLaarhoven8Upper_valid}{CoveringCodes/Database/Sources/VanLaarhoven1989.lean}{635} &
\leanfileat{CoveringCodes/Database/Sources/VanLaarhoven1989.lean}{282} &
Committed data checked by Lean; transcription is untrusted. \\

Trace replay and generated query consistency &
\leanref{UpperTrace.valid}{CoveringCodes/Database/Trace.lean}{108},
\leanref{LowerTrace.valid}{CoveringCodes/Database/Trace.lean}{123},
\leanref{BestBounds.consistent}{CoveringCodes/Database/Defs.lean}{26},
\leanref{bestBounds_consistent}{CoveringCodes/Database/GeneratedAPI.lean}{18} &
\leanfileat{CoveringCodes/Database/Trace.lean}{10},
\leanfileat{CoveringCodes/Database/Defs.lean}{8},
\leanfileat{CoveringCodes/Database/GeneratedAPI.lean}{15},
\leanfileat{CoveringCodes/Database/GeneratedTable.lean}{218} &
Generated traces plus checked replay; generator is unverified. \\
\bottomrule
\end{tabular}
\caption{Map from paper-level claims to checked declarations and artifact files.}
\label{tab:artifact-map}
\end{table*}

\section{Database Closure Diagnostics}\label{app:closure-diagnostics}

As a reproducibility check, the generated table artifact records that
\texttt{lake exe table\_gen} on the current source set generated 52,822
certified table entries and reached a fixpoint after 46 passes.  The executable
reports ``up to 52822 passes'' because this is the coarse termination bound used
by the fixpoint loop; in this run the table stabilizes much earlier.  Here a
fixpoint pass means one complete sweep of the closure loop, not one of the
individual relation relaxations recorded by the trace constructors.

Table~\ref{tab:closure-diagnostics} records the nonzero trace-constructor counts
printed by this run.  These are recursive node counts over all selected
lower-bound and upper-bound traces in the generated table.  They are therefore
diagnostics of the generated proof objects, not independent counts of
mathematical improvements: shared subtraces are counted each time they occur.
The generated output also reports zero-count constructors as audit information;
such rows mean that the constructor is available in the trace language but is
not used by any winning trace for the present source set and parameter range.

\begin{table}[h]
\centering
\footnotesize
\begin{tabular}{lr}
\hline
Trace constructor & Recursive node count \\
\hline
\texttt{UpperTrace.primitive} & 52,822 \\
\texttt{UpperTrace.lengthenFreeN} & 296,174 \\
\texttt{UpperTrace.lengthenDummyN} & 42,090 \\
\texttt{UpperTrace.alphabetExpand} & 402 \\
\texttt{LowerTrace.primitive} & 52,822 \\
\texttt{LowerTrace.radiusBack} & 3,069 \\
\texttt{LowerTrace.lengthBack} & 6,874 \\
\texttt{LowerTrace.lowerBlockGroup} & 34 \\
\hline
\end{tabular}
\caption{Trace-constructor counts reported by \texttt{lake exe table\_gen} for
the current generated database.  Counts are recursive over all nodes in the
selected traces.}
\label{tab:closure-diagnostics}
\end{table}

\section*{Acknowledgements}
\phantomsection
\addcontentsline{toc}{section}{Acknowledgements}

I thank Prof.~Mathar, emeritus professor at RWTH Aachen University, who
introduced me to covering codes in his information theory course.  He was also
an important mentor to me and taught me
mathematical ways of thinking that I have been able to apply throughout my life.

This work was partly funded by the Federal Ministry of Research, Technology and
Space (BMFTR) in Germany under grant number 16KIS2240 of the SUSTAINET-guardian project.

\section*{Author Contributions}
\phantomsection
\addcontentsline{toc}{section}{Author Contributions}

Andreas Florath carried out the Lean formalization, developed the certificate
interfaces and proof-carrying database, extracted and encoded the published
covering-code examples, ran the verification, and wrote the manuscript.

\section*{Use of AI Tools}
\phantomsection
\addcontentsline{toc}{section}{Use of AI Tools}

AI-based tools were used during several stages of this work, including drafting
and revising parts of the manuscript, exploring formulations of explanatory
text, and assisting with software development tasks in the Lean formalization and
supporting tooling.  All mathematical statements, Lean code, proofs, citations,
and manuscript text were reviewed and accepted by the author, who takes
full responsibility for the content of the paper.

\section*{Competing Interests}
\phantomsection
\addcontentsline{toc}{section}{Competing Interests}

The author declares no competing interests.

\onecolumn
\phantomsection
\sloppy
\Urlmuskip=0mu plus 1mu\relax
\printbibliography

\end{document}